\documentclass[journal = chreay, manuscript = article]{achemso}
\usepackage{amsmath}
\usepackage{amssymb}
\usepackage{graphicx}
\usepackage{mathtools}
\usepackage{braket}
\usepackage{hyperref}
\usepackage{bbold}
\usepackage{calrsfs}
\usepackage{dsfont}
\usepackage{xcolor}
\usepackage{subcaption}

\usepackage[scr=boondox]{mathalpha}

\usepackage{soul}

\newcommand{\btau}{\boldsymbol{\tau}}

\def\bra#1{\langle #1 \vert}
\def\ket#1{| #1 \rangle}
\def\braket#1#2{\langle #1 | #2 \rangle}

\makeatletter
\def\@email#1#2{%
 \endgroup
 \patchcmd{\titleblock@produce}
  {\frontmatter@RRAPformat}
  {\frontmatter@RRAPformat{\produce@RRAP{*#1\href{mailto:#2}{#2}}}\frontmatter@RRAPformat}
  {}{}
}%
\makeatother

\renewcommand{\vec}{\boldsymbol}
\title{Collectively-modified inter-molecular electron correlations: The connection of polaritonic chemistry and spin glass physics}

\author{Dominik Sidler}
  \email{dominik.sidler@psi.ch}
  \affiliation{PSI Center for Scientific Computing, Theory, and Data, 5232 Villigen PSI, Switzerland }
 \alsoaffiliation{Max Planck Institute for the Structure and Dynamics of Matter and Center for Free-Electron Laser Science, Luruper Chaussee 149, 22761 Hamburg, Germany}
\alsoaffiliation{The Hamburg Center for Ultrafast Imaging, Luruper Chaussee 149, 22761 Hamburg, Germany}%Lines break automatically or can be forced with \\

\author{Michael Ruggenthaler}
  \affiliation{Max Planck Institute for the Structure and Dynamics of Matter and Center for Free-Electron Laser Science, Luruper Chaussee 149, 22761 Hamburg, Germany}
  \alsoaffiliation{The Hamburg Center for Ultrafast Imaging, Luruper Chaussee 149, 22761 Hamburg, Germany}

\author{Angel Rubio}
  \email{angel.rubio@mpsd.mpg.de}
  \affiliation{Max Planck Institute for the Structure and Dynamics of Matter and Center for Free-Electron Laser Science, Luruper Chaussee 149, 22761 Hamburg, Germany}
    \alsoaffiliation{The Hamburg Center for Ultrafast Imaging, Luruper Chaussee 149, 22761 Hamburg, Germany}
  \alsoaffiliation{Center for Computational Quantum Physics (CCQ) and Initiative for Computational Catalysis (ICC), Flatiron Institute, 162 5th Avenue, New York, NY 10010, USA}
\begin{document}

%%%%%%%%%%%%%%%%%%%%%%%%%%%%%%%%%%%%%%%%%%%%%%%%%%%%%%%%%%%%%%%%%%
%                            Abstract                            %
%%%%%%%%%%%%%%%%%%%%%%%%%%%%%%%%%%%%%%%%%%%%%%%%%%%%%%%%%%%%%%%%%%
\begin{abstract}
%Polaritonic chemistry has attained a rapidly growing interest over the past years, thanks to seminal experimental results demonstrating that site / bond selective chemistry at room temperature is within reach by  collectively strong coupling to the vacuum field fluctuations of optical cavities. However, despite these impressive experimental advances, the fundamental theoretical mechanisms remain elusive. In the following focus review, we uncover a fundamental theoretical connection between the two disconnected research areas \textit{polaritonic chemistry} and the field of \textit{spin glasses}, and explore its far reaching implications on the theoretical understanding of polaritonic chemistry. In particular, we demonstrate an exact mapping of the dressed  electronic-structure problem under collective vibrational strong coupling to the quintessential Sherrington-Kirkpatrick model of a spin glass. The mapping reveals a collectively induced instability criterion  for the dressed  electronic-structure (spontaneous replica symmetry breaking), which could be the so far missing seed that enables the (dynamic) buildup of significant (local) chemical modifications. Furthermore, 
Polaritonic chemistry has garnered increasing attention in recent years due to pioneering experimental results, which show that site- and bond-selective chemistry at room temperature is achievable through strong collective coupling to field fluctuations in optical cavities. Despite these notable experimental strides, the underlying theoretical mechanisms remain unclear. In this focus review, we highlight a fundamental theoretical link between the seemingly unrelated fields of polaritonic chemistry and spin glasses, exploring its profound implications for the theoretical framework of polaritonic chemistry. Specifically, we present a mapping of the dressed many-molecules electronic-structure problem under collective vibrational strong coupling to the spherical Sherrington-Kirkpatrick (SSK) model of a spin glass. This mapping uncovers a collectively induced instability of the intermolecular electron correlations, which could provide the long sought-after seed for significant local chemical modifications in polaritonic chemistry. Overall,  the qualitative predictions made from the SSK model (e.g., dispersion effects, phase transitions, differently modified bulk and rare event properties, heating,...) agree well with available experimental observations.  
Our connection paves the way to incorporate, adjust and probe numerous spin glass concepts in polaritonic chemistry, such as modified fluctuation-dissipation relations, (non-equilibrium) aging dynamics, time-reversal symmetry breaking or stochastic resonances. Ultimately, the connection also offers fresh insights into the applicability of spin glass theory beyond condensed matter systems suggesting novel theoretical directions such as spin glasses with explicitly time-dependent (random) interactions.
\end{abstract}

\tableofcontents

%\date{\today}

\maketitle
 
%%%%%%%%%%%%%%%%%%%%%%%%%%%%%%%%%%%%%%%%%%%%%%%%%%%%%%%%%%%%%%%%%%
%                        Introduction                            %
%%%%%%%%%%%%%%%%%%%%%%%%%%%%%%%%%%%%%%%%%%%%%%%%%%%%%%%%%%%%%%%%%%
%\begin{widetext}

\section{Introduction}

It is well established that molecular properties and chemical reactions can be influenced by light. Femtochemistry~\cite{zewail_laser_1988,zewail_femtochemistry_2000} and coherent control using ultra-short and high-power lasers attest to this. So on a first glance it may seem straightforward to reach for a similar outcome by using optical cavities instead of laser driving, which has established the emergent field of \textit{polaritonic or QED chemistry} (see Fig.~\ref{fig:cavity}).~\cite{ebbesen_hybrid_2016,ruggenthaler_quantum-electrodynamical_2018,sidler_perspective_2022,fregoni_theoretical_2022, ebbesen_introduction_2023, ruggenthaler_understanding_2023, bhuyan_rise_2023, hirai_molecular_2023,simpkins_control_2023,mandal_theoretical_2023,xiang_molecular_2024} There are, however, a few very important differences that make polaritonic chemistry distinct and unique. Firstly, when using lasers one tries to achieve site-selective chemistry usually with coherent, i.e., classical, light fields. In polaritonic chemistry, usually a much smaller number of photons couples and their quantum nature becomes important.~\cite{ebbesen_hybrid_2016,garcia-vidal_manipulating_2021} Secondly, in many cases the electromagnetic field inside an optical cavity is zero, i.e., the cavity is not pumped externally, such that only the strong coupling to the quantum and thermal fluctuations lead to modifications.~\cite{ebbesen_hybrid_2016,garcia-vidal_manipulating_2021} Thirdly, and most importantly for this perspective, in most cases the coupling to a single molecule is small, but non-zero even in the thermodynamic limit.~\cite{ebbesen_hybrid_2016,flick_atoms_2017,svendsen_theory_2023} Therefore, only the macroscopic ensemble of molecules \textit{a priori} couples strongly to the photon-field fluctuations. This \textit{collective-coupling regime} leads to a seemingly paradoxical situation: While chemical reactions and the properties of individual molecules are usually considered local in time and space, this traditional view is challenged in polaritonic chemistry due to strong feedback effects between the microscopic properties and the macroscopic behavior of the ensemble. The unique nature and origin of those feedback effects bridging different scales in time and space will the main focus of the present review. 
\\
\\
Connecting the different scales and isolating the physically relevant mechanism poses a formidable theoretical challenge, which has so far not been resolved satisfactorily due to its complexity (presumably its off-equilibrium glassy nature).~\cite{sidler_polaritonic_2021,sidler_perspective_2022,schnappinger_cavity_2023,sidler_numerically_2023,sidler_unraveling_2024}  Particularly, the origin of why in some molecular ensembles chemical reactions change \cite{hutchison_modifying_2012,thomas_groundstate_2016,lather_cavity_2019,hirai_modulation_2020,fukushima_inherent_2022,ahn_modification_2023,patrahau_direct_2024} while in others under similar conditions no effect is observed,\cite{fidler_ultrafast_2023,chen_exploring_2024} remains elusive. In this article we address the scaling conundrum by using established methods from spin glass physics. Those concepts provide a partial answer and highlight novel ways forward to understand how the macroscopic behavior of an ensemble of molecules can act back on its individual constituents with the help of the electromagnetic modes of an optical cavity. While, as we will detail in this manuscript, there are many intricacies that need more theoretical and experimental investigations, eventually a relatively simple picture will emerge thanks to the established theoretical concepts  of spin glasses. By borrowing ideas from this mature research  discipline and applying them to the situation of collective vibrational strong coupling (VSC, i.e., the cavity is resonant to some vibrational degrees of freedom), we will see that \textit{intermolecular electron correlations} can possess spin glass features under certain conditions. %However, since the overall macroscopic polarization of the ensemble is zero, these synchronized polarizations form intricate patterns that are (dynamically) frustrated.~\cite{mezard_spin_1987}  That is, if the dynamic polarization of one molecule is flipped (or asynchronous), all the others need to react in a concerted synchronized way. In the following, we will call this situation a \textit{polarization glass}.~\cite{sidler_unraveling_2024}
An analogy to the spherical Sherrington-Kirkpatrick (SSK) model  suggest not only a collective \textit{spin glass phase transition}, but it also alters the local fluctuations (rare events) as well as the dynamics of the dressed  electronic-structure  problem. Resulting chemical consequences seem to agree with various experimental data. Moreover, the explicit time-dependence  of the electron correlations (\textit{non-equilibrium aging dynamics}) are expected to act back on all the other degrees of freedom such that it provides a seed to trigger chemically relevant stochastic resonance effects for the (ro)-vibrational degrees of freedom.\cite{sidler_connection_2024} %which together can affect reaction rates (rare events that depend on the details of the thermal ensemble and the local distribution of the energy in the ensemble) and other molecular properties. 
Indeed, based on the analogy with the spin glass, one expects non-trivial local and collective off-equilibrium effects even in a global thermal ensemble. Overall, the presented theoretical framework can provide an avenue for numerous future theoretical and experimental developments in the field of QED chemistry. 
\\
\\
In the following we will first discuss the theoretical setting for describing VSC in the collective regime and discuss how the cavity modifies statistical and thermodynamical considerations (Sec.~\ref{sec:pauli}). Then we show how \textit{intermolecular} correlations lead to a connection to spin glass theory (Sec.~\ref{sec:SSKmodelemerges}). We then discuss some generic properties of spin glasses (Sec.~\ref{sec:spin_glass}), before highlighting potential consequences for polaritonic chemistry under collective VSC and comparing them to various experiments (Sec.~\ref{sec:consequences}). We briefly summarize our findings and future research perspectives in Sec.~\ref{sec:conclusion}.

\section{Pauli-Fierz ab initio theory\label{sec:pauli}}

\begin{figure}
     %\begin{subfigure}[b]{0.5\textwidth}
         \centering
         \includegraphics[width=0.5\textwidth]{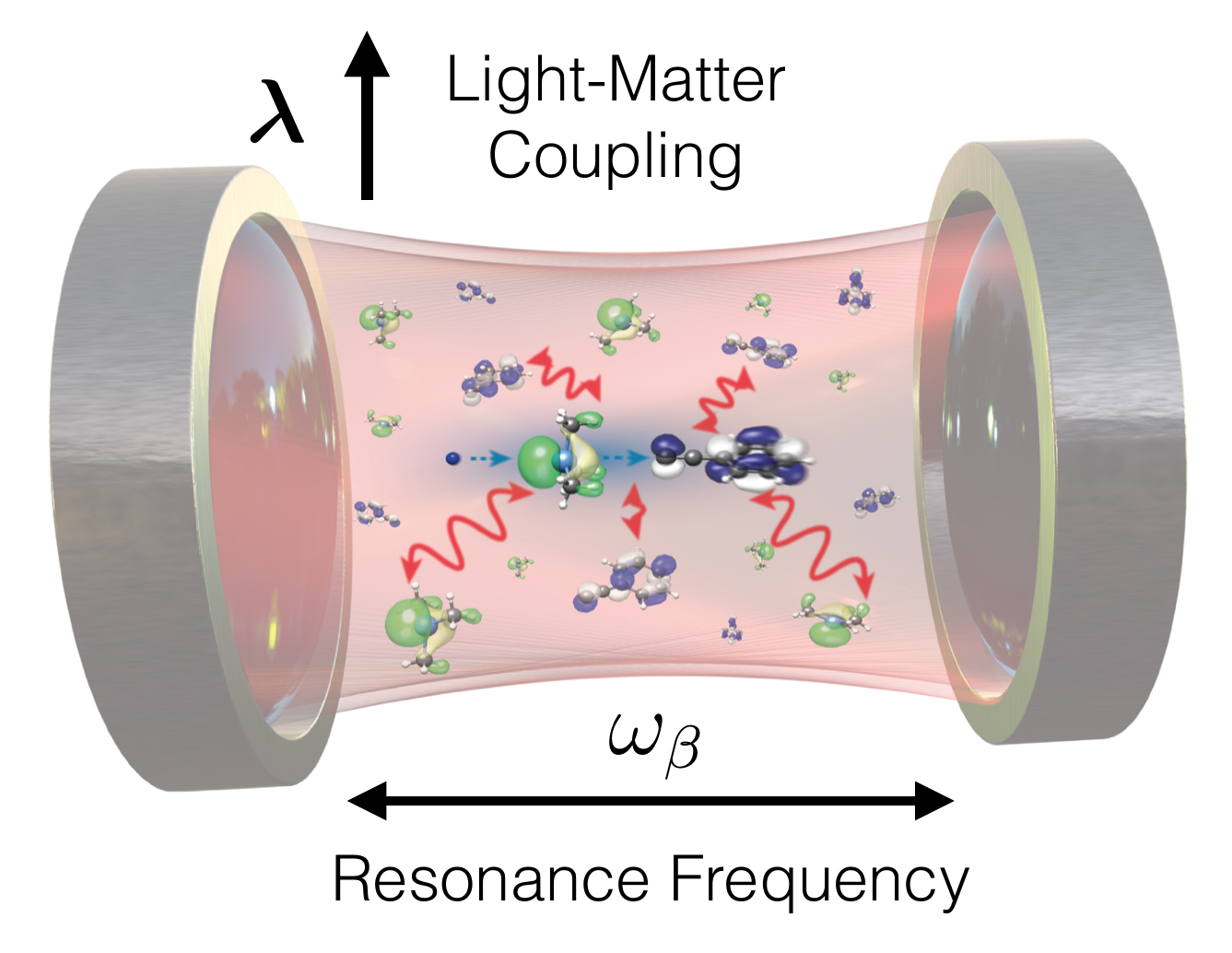}
     %\end{subfigure}
     \caption{Sketch of a molecular ensemble under vibrational strong coupling (VSC) in a Fabry-P\'erot cavity. The distance between the reflective mirrors is inversely proportional the resonance frequency $\omega_{\beta}$, i.e., which photon modes are enhanced due to the standing-wave conditions, and together with the finesse of the mirrors this dictates the single-particle light-matter coupling strength $\lambda$ (see also Eq.~\eqref{eq:lambda-def}).~ \cite{sidler_perspective_2022} }
     \label{fig:cavity}
\end{figure}

As a starting point to describe an ensemble of molecules coupled to an optical cavity one usually employs the Pauli-Fierz theory, which provides a rigorous and non-perturbative theoretical framework to describe the coupling of non-relativistic quantized matter and the quantized light field in an optical cavity (see Fig.~\ref{fig:cavity} for a paradigmatic setup).~\cite{spohn_dynamics_2023, ruggenthaler_understanding_2023} By solving the corresponding Schr\"odinger-type equation, even strongly coupled light and matter can be accurately described on the atomistic scale. In the case that the enhanced light-modes of the optical cavity have a wavelength much larger than the molecular systems, we can employ the long-wavelength and the few-mode approximation~\cite{ruggenthaler_understanding_2023}, such that in length-gauge the Pauli-Fierz Hamiltonian takes the form
\begin{equation}
    \hat{H} = \hat{H}^{\rm m}
    +\frac{1}{2}\bigg[\hat{p}_\beta^2+\omega_\beta^2\Big(\hat{q}_\beta-\frac{\hat{X}+\hat{x}}{\omega_\beta}\Big)^2\bigg].
    \label{eq:pf_dip_h}
\end{equation}
For simplicity we have chosen a single-effective cavity mode $\beta$, e.g., of a perfect (dissipationless) Fabry-P\'erot cavity. Notice that more evolved cavity setups can be designed that, e.g., allow for higher mode-volume confinements (such as in plasmonic or micro cavities).\cite{hugall_plasmonic_2018,vahala_optical_2003,torma_strong_2015,chang_colloquium_2018,skolnick_strong_1998}  Moreover, cavity leakage effects of the mirrors are a priori not captured by Eq.~\eqref{eq:pf_dip_h}. However, those could be accounted for by considering multiple modes (broadening) and thus introducing a finite linewidth (lifetime) which leads in the continuum limit to the imaginary part of the dielectric response of the mirrors.\cite{flick_lightmatter_2019,svendsen_ab_2024}  The more general minimal-coupling Pauli-Fierz framework is discussed in, for example, Ref.~\citenum{ruggenthaler_understanding_2023}, but we do not expect that this more intricate description will qualitatively change the results in the following. 

In Eq.~\eqref{eq:pf_dip_h} the free-space matter Hamiltonian is defined as $\hat{H}^{\rm m}$, which accounts for the quantized nuclei and electrons of the molecules. Thus $\hat{H}^{\rm m}$ describes the highly non-trivial free-space matter problem, which has been the focus of quantum-chemical methods over many decades. The second term describes the coupling of the matter to the quantized displacement field operator $\hat{q}_\beta$, with the conjugate photon operator defined as $\hat{p}_\beta$. The matter polarization operators for $N$ molecules with $N_n$ nuclei and $N_e$ electrons are given as
$\hat{X}:=\boldsymbol{\lambda}\cdot\sum_{i=1}^N \sum_{n=1}^{N_n} Z_n \boldsymbol{\hat{R}}_{in}$ and $ \hat{x}:=-\boldsymbol{\lambda}\cdot\sum_{i=1}^{N} \sum_{n=1}^{N_e}  Z_e \boldsymbol{\hat{r}}_{i n}$, 
where the nuclear and electronic total transition dipole moments, respectively, are coupled via $\boldsymbol{\lambda}$ to the effective displacement field mode of frequency $\omega_\beta$.  The vectorial photon-matter coupling $\boldsymbol{\lambda} = \boldsymbol{\varepsilon} \lambda$ depends on the mode polarization vector $\boldsymbol{\varepsilon}$ and the coupling constant~\cite{ruggenthaler_understanding_2023}
\begin{equation}\label{eq:lambda-def}
    \lambda = \sqrt{\frac{e^2}{\mathcal{V} \varepsilon_0}},
\end{equation}
where $\mathcal{V}$ corresponds to the effective mode volume. This effective mode volume can be connected to properties of the Fabry-P\'erot cavity and scales roughly as $L^3 \mathcal{F}$, where $\mathcal{F}$ is the finesse of the cavity. A more elaborate discussion reveals that the effective mode volume leads to a finite light-matter coupling even in the macroscopic limit.~\cite{svendsen_theory_2023}. It is important to highlight two aspects of the Pauli-Fierz theory in the length gauge and the few-mode approximation. Firstly, that we have an equilibrium solution of the coupled system is due to the fact that the light-matter coupling term in Eq.~\eqref{eq:pf_dip_h} is quadratic and thus the Hamiltonian is bounded from below. Of specific importance for the stability of the coupled system is the term $(\hat{X} + \hat{x})^2$, which is quadratic in the coupling strength $\lambda$.~\cite{rokaj_lightmatter_2018,schafer_relevance_2020} This term is called dipole self-energy or self-polarization term in the literature. Secondly, for the proper free-space continuum limit it is important to subtract the free-space contributions from the effective-mode theory. Otherwise, one would double-count the interaction with the free-space modes that are captured by working with the observable masses of the charged particles.~\cite{svendsen_theory_2023} Thus $\lambda =0$ means no cavity, and for any Fabry-P\'erot cavity there is a finite mode volume $\mathcal{V}$ which implies $\lambda > 0$ and also dictates the maximal amount of molecules that can in fact be coherently coupled for a given molecular density. This aspect will become important later when we discuss the scaling behavior in QED chemistry (see Sec.~\ref{sec:instability}). 
%\revDS{Ruggi: Mention bounded from below, i.e. minimum free space $\lambda$. What is minimum $\lambda$? Can it be expressed in terms of natural constants?}

At this point it is important to highlight again that in most experiments in polaritonic chemistry the effective mode volume is large, since many molecules are coherently coupled, and $\lambda \ll 1$. Based on this small prefactor it is then often argued that no effect for molecular systems should be observed in a dark cavity. We will, however, not apriori discard the coupling terms. Most importantly because although the prefactor $\lambda$ might be small, the quadratic coupling term formally scales as $N^2$ and hence for $N \gg 1$ this scaling can potentially balance this small prefactor and eventually give rise to a quantitative effect at the single molecular level. Moreover, because in the following we focus on VSC, which implies that the cavity frequency $\omega_\beta$ is tuned on the vibrational excitations rather than the energetically higher-lying electronic excitions, the cavity Born-Oppenheimer partitioning is a natural and effective choice to proceed.~\cite{flick_atoms_2017,flick_cavity_2017,ruggenthaler_understanding_2023} Thus in a Born-Huang expansion the total wave function is partitioned by grouping the nuclear and displacement field degrees and the electronic wave function is treated as a conditional wave function that depends on the nuclear and displacement coordinates. This allows to write the Hamiltonian for the electronic part of the coupled problem as 
\begin{align}
 \hat{H}^{\rm e}(\boldsymbol{R},q_\beta):=&H^{\rm m,e}(\boldsymbol{R})
 +\bigg(\frac{1}{2} \hat{x}^2+\hat{x} X-\omega_\beta  \hat{x} q_\beta\bigg),\label{eq:Helectron}
\end{align}
which parametrically depends on all the nuclei positions 
\begin{align}
\boldsymbol{R} := [\boldsymbol{R}_1 =(\boldsymbol{R}_{11}, \dots, \boldsymbol{R}_{1 N_n}), \dots, \boldsymbol{R}_N = (\boldsymbol{R}_{N1}, \dots, \boldsymbol{R}_{N N_n})]
\end{align} 
and displacement photon field coordinates, written compactly as $(\boldsymbol{R},q_\beta)$. The free-space electronic-structure problem is given by $H^{\rm m,e}(\boldsymbol{R})$.
Notice that if we would keep all non-adiabatic couplings in the Born-Huang expansion no approximation has been made so far. Only in a next stage, different levels of approximations are introduce to reduce the computational complexity of the fully quantized problem given in Eq.~\eqref{eq:pf_dip_h}, see e.g. Refs.~\citenum{flick_cavity_2017,schafer_ab_2018}.
\\
\\
Throughout this work, we are mainly interested in the physical properties of the cavity-mediated  electronic-structure given in Eq.~\eqref{eq:Helectron}. For this reason, we proceed by applying the classical (for the nuclear and displacement degrees of freedom) cavity Born-Oppenheimer approximation on the coupled nuclear-photon problem. This allows for a computationally efficient determination of reasonable parameters $(\boldsymbol{R},q_\beta)$ that enter the dressed electronic-structure problem. In more detail, nuclei and (effective~\cite{svendsen_theory_2023}) displacement field evolve on the dressed ground-state electronic potential-energy surface according to the classical Hamiltonian dynamics of,~\cite{sidler_perspective_2022, sidler_unraveling_2024,hoffmann_capturing_2019,chen_ehrenfestr_2019-1,li_mixed_2018,chen_ehrenfestr_2019,hoffmann_benchmarking_2019,fregoni_photochemistry_2020,li_collective_2021}
\begin{align}
   H^{\mathrm{npt}} :=  H^{\rm m,n}(\boldsymbol{R})
    +\frac{p_\beta^2}{2}
    +\frac{\omega_\beta^2}{2}\Big(q_\beta-\frac{X}{\omega_\beta} \Big)^2 +\bra{\Psi_0}\hat{H}^\mathrm{e}(\boldsymbol{R},q_\beta)\ket{\Psi_0}.\label{eq:class_ham}
\end{align}
The classical cavity Born-Oppenheimer approximation implies that any quantum and non-adiabatic effects of the nuclear structure are subsequently discarded, which allows for ground-state ab initio molecular dynamic implementations. Such a theoretical setup is numerically feasible even for large molecular ensembles $N\gg 1$. We will get back to this adiabatic assumption and discuss its validity in the context of VSC later in Secs.~(\ref{sec:instability},\ref{sec:re_breaking},\ref{sec:offequil}). Notice that assuming a classical displacement field $\boldsymbol{D}=\boldsymbol{\lambda}\omega_\beta q_\beta/4\pi$ does not mean that the transverse electric-field operator $\hat{\boldsymbol{E}}_\perp$ is entirely classical. Indeed, the transverse electric-field operator is given as,~\cite{sidler_unraveling_2024}
\begin{eqnarray}
    \hat{\boldsymbol{E}}_\perp= 4\pi (\boldsymbol{D}-\hat{\boldsymbol{P}}) \label{eq:Etrans}
\end{eqnarray}
within the length-gauge representation used throughout this paper. Thus the electronic part of the macroscopic polarization operator $\hat{\boldsymbol{P}}=\boldsymbol{\lambda}(X+\hat{x})/4\pi$ remains fully quantized and describes the electronically bound photons of the hybrid light-matter states.\cite{rokaj_lightmatter_2018,ruggenthaler_understanding_2023} The quantized nature of $\hat{\boldsymbol{P}}$ will be  an essential ingredient for all the subsequent discussions. A further important point is that the displacement field coordinate couples to the \textit{total} dipole of the molecular ensemble and the coupling scales linear in $\lambda$. Thus the coupling of the displacement field is different from the direct dipole-dipole coupling due to the self-energy term. Notice further that the longitudinal electric fields remain unaffected by the gauge choice, i.e., they correspond to the standard Coulomb interaction terms of the bare matter problem with classical nuclei and quantized electrons. 
\\
\\

%%%%%%%%%%%%%%%%%%%%%%%%%%

\subsection{Cavity Hartree-Fock (cHF) approximation}
Assuming without loss of generality a system of $N$ identical molecules, each one possessing $N_e$ electrons, the cavity-Born-Oppenheimer Hartree-Fock (cBOHF) electronic wavefunction of the $(N \times N_{e})$-electron system is a Slater determinant of mutually orthonormal spin orbitals $\varphi_i$~\cite{szabo_modern_2012}:
\begin{equation}
 \label{eq:sd}
\Psi(\btau_{11},\ldots \btau_{N_e 1}, \ldots, \btau_{N_e N}) = \frac{1}{\sqrt{N_{e}!}} \braket{\btau_{11}, \ldots, \btau_{N_e1}, \ldots \btau_{N_eN}}{\varphi_{11},\ldots, \varphi_{N_e1}, \dots \varphi_{N_e N}}\,.
\end{equation} 
Here, $\btau_i=(\boldsymbol{r}_i \sigma_i)$ is used to denote the complete set of coordinates associated with the $i$-th electron, comprised of the spatial coordinate $\boldsymbol{r}_i$ and a spin coordinate $\sigma_i$. 
Note that the $(N \times N_{e})$-electron system described by $\Psi$ can be a single molecule (if $N=1$) or an ensemble of many molecules (if $N>1$). Thus, it is possible to treat cavity-induced interactions and standard Coulomb interactions in the same way. It is important to highlight that in the ensemble case we have besides intramolecular also intermolecular Coulomb and cavity-induced interactions. 
For the special case of the dilute gas limit, that is the  electronic-structures of different molecules do not overlap, the ensemble Slater determinant may be replaced by a product~\cite{sidler_unraveling_2024} of individual molecular Slater determinants. Note that the displacement coordinate
$q_\beta$ of the electric field mode is treated as a parameter in cBOHF ansatz, analogously to the nuclear coordinates, and thus is not part of the wave function. The resulting energy expression, where we have for the sake of brevity relabeled the electronic and nuclear coordinates as single sums, takes the form 
\begin{eqnarray}
\label{eq:e_cbo}
\bigl\langle  \Psi\big| \hat{H}^e\big|\Psi\bigl\rangle &=& \bigl\langle  \Psi \big| \sum_i^{N_e N}\bigg\{\frac{\hat{\vec{p}}_i^2}{2} -\frac{1}{2}\sum_l^{N_n N} \frac{Z_l}{|\hat{\vec{r}}_i - \vec{R}_l|}+\frac{1}{2}\sum_j^{N_e N}\frac{1}{|\hat{\vec{r}}_i - \hat{\vec{r}}_j|}\bigg\}+
\bigg(\frac{1}{2} \hat{x}^2+\hat{x} X-\omega_\beta  \hat{x} q_\beta\bigg) \big| \Psi  \bigr\rangle \nonumber\\
 &=&
\sum_i^{N_e N}\int d \btau \phi_i^*(\btau)\left(\hat{h}^{\rm m}+\hat{h}^{\rm l}\right)\phi_i(\btau)\nonumber \\
&&+ \frac{1}{2}\sum_i^{N_e N} \sum_{j}^{N_e N} \int d \btau_1 \int d \btau_2 \phi_i^*(\btau_1) \phi_j^*(\btau_2)\frac{1}{|\hat{\vec{r}}_1 - \hat{\vec{r}}_2|}\phi_i(\btau_1)\phi_j(\btau_2)\nonumber\\
&&- \frac{1}{2}\sum_i^{N_e N}\sum_{j}^{N_e N}\int d \btau_1 \int d \btau_2 \phi_i^*(\btau_1) \phi_j^*(\btau_2)\frac{1}{|\hat{\vec{r}}_1 - \hat{\vec{r}}_2|}\phi_i(\btau_2)\phi_j(\btau_1)\nonumber\\
&&+ \frac{1}{2} \sum_i^{N_e N}\sum_{j}^{N_e N}\int d \btau  \phi_i^*(\btau)\vec{\lambda}\cdot\hat{\vec{r}} \phi_i(\btau) \int d \btau \phi_j^*(\btau)\vec{\lambda}\cdot\hat{\vec{r}}\phi_j(\btau)\nonumber\\
&&-\frac{1}{2} \sum_i^{N_e N}\sum_{j}^{N_e N}\bigg|\int d \btau  \phi_i^*(\btau)\vec{\lambda}\cdot\hat{\vec{r}} \phi_j(\btau) \bigg|^2\nonumber \\
%%%%%%%%%%
&=& \underbrace{\big\langle \hat{h}^{\rm m}\big\rangle}_{=h^{\rm m}}+\underbrace{\big\langle\hat{h}^{\rm l} \big\rangle}_{=h^{\rm l}}+\underbrace{\big\langle\hat{J}_{\rm coul}\big\rangle}_{=J_{\rm coul}}+
  \underbrace{\big\langle \hat{K}_{\rm coul}\big\rangle}_{=K_{\rm coul}}+\underbrace{\big\langle\hat{J}_{\rm DSE}\big\rangle}_{=J_{\rm DSE}}+
  \underbrace{\big\langle\hat{K}_{\rm DSE}\big\rangle}_{=K_{\rm DSE}}\label{eq:SCF_slater}
\end{eqnarray}
where we used $\hat{h}^{\rm m}=\tfrac{\hat{\vec{p}}^2}{2} -\frac{1}{2}\sum_l^{N_n N} Z_{l}/(|\hat{\vec{r}} - \vec{R}_l|)$ and $\hat{h}^{\rm l}= \tfrac{1}{2}(Z_{e} \boldsymbol{\lambda}\cdot\hat{\boldsymbol{r}})^2- Z_{e} X\boldsymbol{\lambda}\cdot\hat{\boldsymbol{r}} +\omega_\beta q_\beta Z_{e} \boldsymbol{\lambda}\cdot\hat{\boldsymbol{r}} $. The two-electron integrals can be split into four different contributions: The Coulomb Hartree integral $J_{\rm coul}$, the Coulomb exchange integral $K_{\rm coul}$, the dipole-self energy (DSE) Hartree integral $J_{\rm DSE} $, and the DSE exchange integral $K_{\rm DSE}$. The standard variational procedure to search for the determinant $\ket{\Psi}$ with minimal energy $E_0=\bra{\Psi_0}\hat{H}^e\ket{\Psi_0}$ is by variation of the spin orbitals $\phi_i$ such that $\delta \bra{\Psi}\hat{H}^e\ket{\Psi}=0$, together with an orthogonality constraint $\braket{\phi_i}{\phi_j}=\delta_{ij}$. This results in a non-linear eigenvalue problem for molecular orbitals that can be written in a compact form as
\begin{eqnarray}
   \underbrace{\left[\hat{h}^{\rm m}+ \hat{h}^{\rm l}+\hat{J}_{\rm coul}-\hat{K}_{\rm coul}+\hat{J}_{\rm DSE}-\hat{K}_{\rm DSE}\right]}_{=\hat{F}}\phi_k(\btau)=
    \epsilon_k\phi_k(\btau)\label{eq:cHF}
\end{eqnarray}
Here the  Fock operator $\hat{F}$ depends on the spin-orbitals $\phi_k(\btau)$ and is thus non-linear. This eigenvalue problem is usually computed iteratively. 
\\
\\
Before we move on, we make an important observation about the accuracy of the Hartree-Fock method for a large ensemble of molecules or atoms. In the case of zero light-matter coupling, i.e., outside a cavity, it can be shown that the Hartree-Fock method becomes asymptotically exact as we approach the bulk/thermodynamic $N \rightarrow \infty$ limit~\cite{bach_error_1992,lieb_stability_1976,lieb_thomas-fermi_1981}.%[https://link.springer.com/article/10.1007/BF02097241],~[https://journals.aps.org/rmp/abstract/10.1103/RevModPhys.48.553],~[https://journals.aps.org/rmp/abstract/10.1103/RevModPhys.53.603].
The reason for this is that the ground-state energy of the total ensemble scales as $N_{\rm tot}^{7/3}$, with $N_{\rm tot} = N N_e$ for charge neutral systems, while the difference between the exact and the Hartree-Fock solution scales as $N_{\rm tot}^{11/5}$. We therefore have that $\lim_{N_{\rm tot} \rightarrow \infty} (E_{\rm HF}[N_{\rm tot}] - E_{\rm exact}[N_{\rm tot}])/E_{\rm exact}[N_{\rm tot}] \rightarrow 0$. That is, the Hartree-Fock error compared to the total energy of the system vanishes. Such results are commonly used in the theory of phase transitions, where it is usually assumed that mean-field descriptions become asymptotically exact in the thermodynamic limit, i.e., for $N\rightarrow \infty$~\cite{kadanoff_more_2009}. In this context it is also important to highlight that this argument does not imply that \textit{intramolecular} properties of an individual molecule of the ensemble are treated exactly in this limit, but the behavior of the total ensemble is well represented by Hartree-Fock (or even Hartree) theory. This is why a statistical thermodynamic description of the ensemble of molecules becomes accurate~\cite{kadanoff_more_2009}. And when we consider chemical rates, they are as much a property of the individual constituent as of the total ensemble itself~\cite{yip_handbook_2005}. Moreover, we need to highlight that for the behavior of an ensemble of molecules, the ground-state degeneracies and the amount of lowest-lying excited states are decisive~\cite{aizenman_third_1981}. That is, the amount of available states (of the total ensemble) within an energy range set by the temperature, determines its phase and behavior. All of this will now be (at least slightly) modified by the cavity, which breaks the symmetry of free space and imposes a new time and length scale onto the molecular ensemble. Indeed, for the thermodynamic stability of the ensemble of molecules it is important to realize that the Coulomb-interaction gets screened and is short-ranged on an intermolecular scale~\cite{kadanoff_more_2009}, while the DSE terms are long-ranged due to their transverse nature and hence connect many more molecules. Let us elucidate these different points now in the following sections.

\subsection{Dilute gas limit: cavity Hartree equations}

Let us start with the simplest possible case to investigate the difference between free space and a cavity. For this we assume the dilute-gas limit for the corresponding  many-electron wave function $\Psi$, i.e., free-space molecules do not interact with each other such that we can approximately replace $H^{\rm m,e}(\boldsymbol{R}) \rightarrow \sum_{i=1}^N H^{\rm m,e}_i(\boldsymbol{R}_i)$. Under this assumption one can determine the properties of a molecular ensemble from just solving a single representative molecule, which is the focus of usual quantum-chemistry methods~\cite{szabo_modern_2012}. The gaseous ensemble properties are then determined by doing classical, independent statistics on top of the single-molecule spectrum. In this way one can connect to emission and absorption spectra of molecular ensembles. Indeed, these spectra are usually ensemble properties and the different peaks correspond to highly degenerate ensemble states. That is, there are very many combinations of single-molecule excitations that have the same total excitation energy of the ensemble. This simple setting is 
%\revDS{todo: update this subsection. first part is copied from last submission and needs to be updated to connect with previous cHF picture}
%To investigate the physical properties of the dressed electronic problem in Eq.~\eqref{eq:Helectron}, it is convenient to assume the dilute-gas limit for the corresponding  many-electron wave function $\Psi$, i.e., free-space molecules do not interact with each other described by $H^{\rm m,e}(\boldsymbol{R})=\sum_{i=1}^N H^{\rm m,e}_i(\boldsymbol{R}_i)$. This is 
a common choice in the field of polaritonic chemistry, which is applied to a broad range of different situations.~\cite{tavis_exact_1968,galego_cavity-induced_2015,schafer_polaritonic_2022, schnappinger_cavity_2023,sidler_unraveling_2024} Indeed, in principle we can extend this ansatz to molecules in more complex environments, such as in solution. In this case $H_i^{\rm m,e}(\boldsymbol{R}_i)$ does not correspond to the  electronic-structure of a single molecule but to a full solvation shell instead. And then classical, independent statistics are performed on top of this repeating unit.
%\revDS{Probably add more details (later?), why this seems reasonable even in solvation (local different perturbation can also seen as different parameters, which seems not to change too much overall properties, analogy of SK and EA model etc.)} 
Remaining in the simple-dilute gas case, we essentially assume non-overlapping  electronic-structures between the $N$ molecules, which reduces the total electronic wave function $\Psi$ to a simple Hartree product $\Psi = \psi_1 \otimes \psi_2 \otimes \dots \otimes \psi_i \otimes \dots \otimes \psi_{N}$ of $N$ single-molecule electronic wave functions $\psi_i$. The only difference to the free-space case comes in this level of approximation from the long-range $\hat{J}_{\rm DSE}$ term. Performing a minimzation with respect of the individual single-molecule wave functions (which internally still have all the \textit{intramolecular} Coulomb and DSE contributions) are determined by the following $N$ coupled cavity Hartree~(cH) equations,~\cite{sidler_unraveling_2024, schnappinger_cavity_2023}
\begin{align}
%\label{eq:CBO_electronic}
 &\bigg(H^{\rm m,e}_i(\boldsymbol{R}_i)+ \Big(X - q_\beta \omega_{\beta} \!+\!\! \sum_{j \neq i}^{N} \bra{\psi_j}\hat{x}_{j}\ket{\psi_j} \Big) \hat{x}_{i}+
\frac{\hat{x}_{i}^2 }{2}  \bigg) \Psi_i = 	\varepsilon_i \Psi_i. \label{eq:cavhartree}
\end{align}	
Notice that the mean-field (direct product) description implies no quantum entanglement between the molecules. While in principle one could go beyond a mean-field theory, e.g., using coupled cluster, and other more accurate ab-initio methods\cite{haugland_coupled_2020,mordovina_polaritonic_2020,rivera_variational_2019,ahrens_stochastic_2021,schafer_making_2021,sidler_chemistry_2020,flick_lightmatter_2019,yang_quantum-electrodynamical_2021,welakuh_frequency-dependent_2022} for the collective electronic-structure problem, it comes at the cost of increasing the computational load considerably. However, as discussed above, if we are interested on the macroscopic scale the mean-field treatment (at least for the free-space case) becomes asymptotically exact, and we can expect reasonable results already on this level of mean-field theory.~\cite{kadanoff_more_2009,golse_mean_2016} 
%However, this additional complexity might not be necessary for large molecular ensembles under collective strong coupling, since it has been shown for similar situations that a mean-field treatment becomes exact in the large $N$-limit.~\cite{kadanoff_more_2009,golse_mean_2016} 
However, the rigorous mathematical analysis of this aspect when coupled to a multimode cavity remains an open research question for the moment. Nevertheless, the subsequently developed connection to a classical spin glass is expected to capture the most relevant physical mechanism of the dressed \textit{intermolecular} electronic-structure problem under VSC. Certainly, the \textit{intramolecular} electronic-structure problem (bare matter) requires the inclusion of exchange and correlation terms to be chemically accurate. However, since we do not specify the level of theory on the \textit{intramolecular} scale in Eq.~\eqref{eq:cavhartree}, this is implicitly enabled by the cavity Hartree equations.
\\
\\
To determine the dressed many-molecule ground state, the $N$ coupled cavity Hartree equations need to be solved iteratively until convergence. One immediately notices that the recursive dependency on $\sum_{j \neq i}^{N} \bra{\psi_j}\hat{x}_{j}\ket{\psi_j} $ may introduce a significant (non-perturbative) modification of the  electronic-structure, even for small coupling constants $\lambda$.~\cite{kadanoff_more_2009} This terms originates from the quadratic interaction term $\hat{x}^2$ in Eq.~\eqref{eq:pf_dip_h} that has the formal scaling of $N^2$. Notice that the length-gauge representation is convenient to uncover this fundamental all-to-all interaction term. However, it is a gauge-independent feature, which is also present in any other gauge choice (e.g., also in the common velocity gauge).~\cite{schafer_making_2021,lu_electron-photon_2024} It is the main difference to a free-space ensemble, where the macroscopic state (which merely statistically explores all single-molecule configuartions) has no influence on the individual molecules. In the following, we will first investigate the consequences of this novel long-range all-to-all interaction on the individual constituents. Afterwards, we complement our picture by including cavity-mediated electron correlations in denser molecular ensembles under VSC.

\subsection{Effective-electron approximation}

To better understand the impact of the DSE Hartree interaction $\hat{J}_{\rm DSE}$ on the \textit{intermolecular}  electronic-structure of the molecular ensemble in a cavity, we additionally assume that every molecule has only one effective electron, i.e., setting $N_e = N$. This is a simple way of capturing the (electronic) polarizability of individual molecules, which allows to focus on the \textit{intermolecular} properties of our ensemble. Those are assumed to not depend critically on the microscopic \textit{intramolecular} details. 
%and the physics of our problem is entirely determined by solving~\cite{sidler_unraveling_2024, schnappinger_cavity_2023}
%\begin{eqnarray}
% \underset{\boldsymbol{c}}{\rm min}\bigl\langle  \Psi\big| \hat{H}^e\big|\Psi\bigl\rangle\overset{N_e=N,\ \rm dilute\ gas}{\rightarrow}\min_{\bm{c}}\big\langle \hat{h}^{\rm m}+\hat{h}^{\rm l}+\hat{J}_{\rm DSE}
%  \big\rangle_{\bm{c}}\label{eq:cHmin}
%\label{eq:CBO_electronic}
 %&\bigg(H^{\rm m,e}_i(\boldsymbol{R}_i)+ \Big(X - q_\beta \omega_{\beta} \!+\!\! \sum_{j \neq i}^{N} \bra{\psi_j}\hat{x}_{j}\ket{\psi_j} \Big) \hat{x}_{i}+
%\frac{\hat{x}_{i}^2 }{2}  \bigg) \Psi_i = 	\varepsilon_i \Psi_i. \label{eq:cavhartree}
%\end{eqnarray}
%The consequences of the cavity Hartree equations under collective strong coupling \revDS{mention explicitly in abstract that we care about collective VSC}conditions, i.e., Eq. \eqref{eq:cHmin}, have recently been discussed in Refs. \cite{sidler_unraveling_2024,horak_analytic_2024} numerically as well as analytically.
%\subsubsection{Analytic results}
%\\

For one-dimensional and harmonic models of molecules, the  self-consistent cavity Hartree problem in Eq. \eqref{eq:cavhartree} can be solved analytically for arbitrarily many molecules $N$, as recently demonstrated in Ref.~\cite{horak_analytic_2024}. In this simplified setting, we find an analytic expression for the feedback of the ensemble of molecules on the cavity mode, i.e., due to the change of refractive index~\cite{fiechter_understanding_2024,horak_analytic_2024}, in form of a renormalization of the cavity frequency $\tilde{\omega}_\beta$ (red-shift) as
\begin{eqnarray}\label{eq:redshift}
    \tilde{\omega}_\beta^2&=&\gamma^2 \omega_\beta^2\\
    \gamma^2&=& \frac{1}{1+\lambda^2 N\alpha_i}\leq 1.
\end{eqnarray}
Here $\alpha_i$ corresponds to the bare matter polarizability of a single molecule and the redshift parameter $\gamma$ scales with the system size $N$ and thus depends on the collective coupling strength. The most notable analytic result is a cavity-modified local (!) molecular polarizability 
\begin{eqnarray}
   \tilde{\alpha}_i = \gamma^2 \alpha_i < \alpha_i.
   \end{eqnarray}
   Here the local molecular polarizability is defined by applying an external electric field $E_\mathrm{ext}$ to the full ensemble as~\cite{horak_analytic_2024}
   \begin{eqnarray}
   \tilde{\alpha}_i = Z_e\frac{\partial \langle  r_i\rangle}{\partial E_\mathrm{ext}} \ \mathrm{with } \ \hat{H}_{\rm tot}=\hat{H}^e - Z_e\sum_i^N\hat{r}_i E_\mathrm{ext}.
\end{eqnarray}
Recent ab-initio calculation indicate that local polarizability modifications also exist for real molecular systems under collective strong coupling.\cite{schnappinger_molecular_2025}
In contrast, treating the external electric field with perturbation theory on the single-molecule level, where $\lambda$ seems a small parameter, the local polarizability of the molecule remains the free-space one, i.e., 
\begin{eqnarray}
   \tilde{\alpha}_i^{\rm pert}  \overset{N\gg 1}{\rightarrow} \alpha_i.
   \end{eqnarray}
It is easy to understand statistically why the usual trick of reducing to a single-molecule description, as is implicitly assumed in free space, does not work. The Hartree DSE term makes all the molecules statistically dependent, and hence we need to consider all possible cavity-mediated interactions between the molecules in the ensemble. From an energy perspective, we can equivalently say that while the prefactor $\lambda^2$ is small in $\langle\hat{J}_{\rm DSE}\rangle$, we have $N^2$ contributions that potentially have to be taken into account. This simple example shows that a self-consistent treatment of the full ensemble can become important to determine local properties under VSC.
%This demonstrates, that the self-consistent treatment of the response of the cavity-modified  electronic-structure can be essential to capture all relevant physical mechanism and particularly local effects can emmerge from collective strong coupling effects in large molecular ensembles. 

\subsubsection{Polarization glass}\label{subsubsec:polarizationglass}
%\revDS{Todo update the following subsection, which was taken from first submission:}

%\revDS{do we have data that show that polarization is averaged to zero on a single molecule? i.e. $\Delta \mu=0$, but $\Delta Var\mu \neq0$? This we cannot represent in two level system, but important to connect to Ebbesen results. Can we determine it from the aligned molecular results? Howver only possible if the initial state of the shin-metiu molecules is not equally distributed between state A and B! Check code. Otherwise polarization naturally cancels globally again. States A and B are randomly initialized! Add distribution plot $P(\Delta \mu)$ for shin-metiu! Necessary to connect to results of Ebbesen, since 2 level system cannot account for this effect! If there is net polarization, probably interpret it with different entities (e.g. solvent rearrangement changes polarization locally, but does not appear in NMR spectra).
%\subsubsection{Numerical polarization glass}

\begin{figure}
     %\begin{subfigure}[b]{0.5\textwidth}
         \centering
         \includegraphics[width=0.5\textwidth]{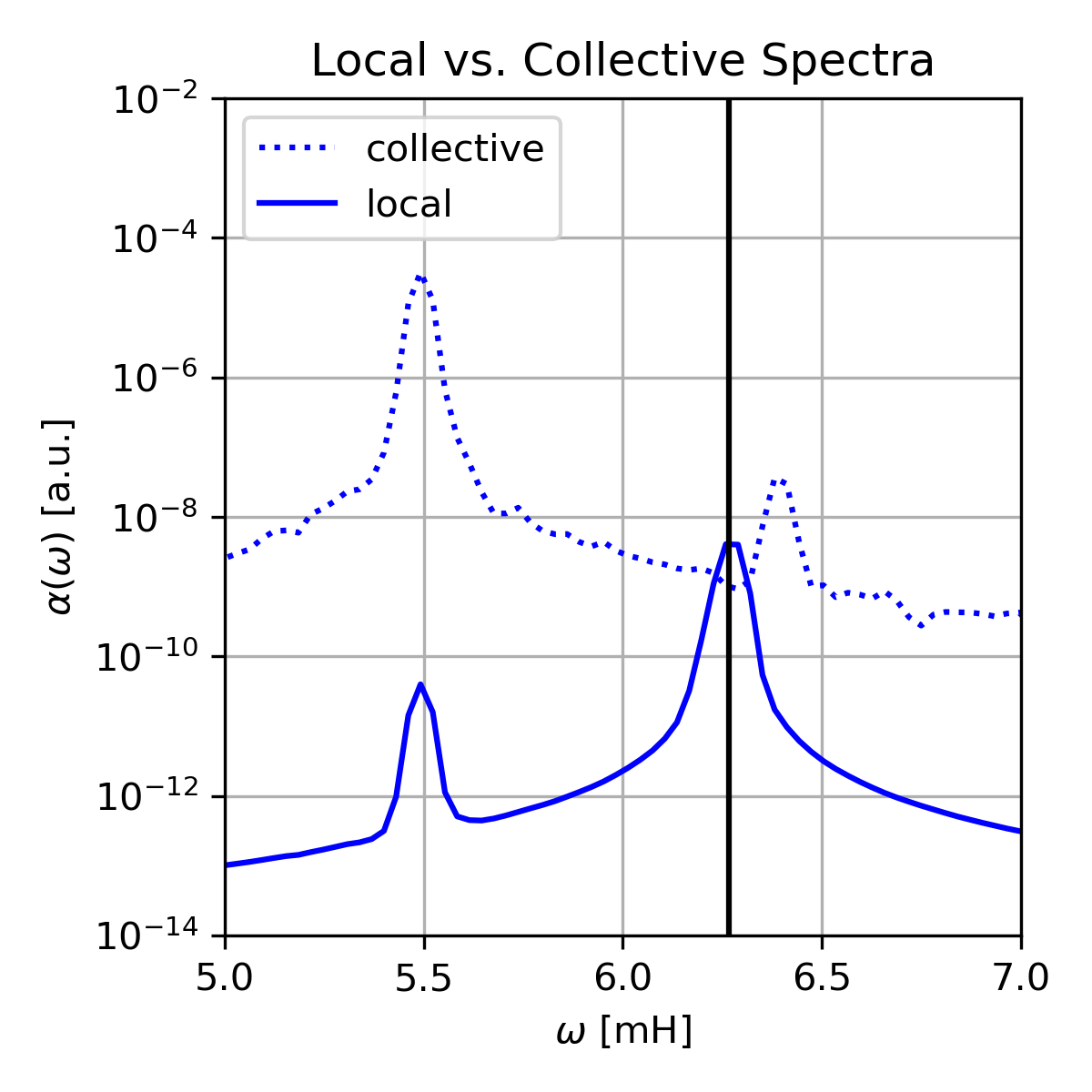}
     %\end{subfigure}
     \caption{Collective (dotted) vs. local (bold) Rabi splitting for (N=900) aligned Shin-Metiu molecules under VSC, taken from Ref.~\citenum{sidler_unraveling_2024}. The local upper polariton is hidden in the broadening of the dark states, which occur at the bare cavity frequency (vertical black line). The asymmetry of the collective Rabi splitting with respect to the bare cavity frequency is a consequence of the red-shift that is caused by the polarizability of the medium, i.e., due to the dipole self-interaction term in the Hamiltonian. }
     \label{fig:splitting}
\end{figure}

\begin{figure}
         % \begin{subfigure}[b]{0.45\textwidth}
         \centering
         \includegraphics[width=0.5\textwidth]{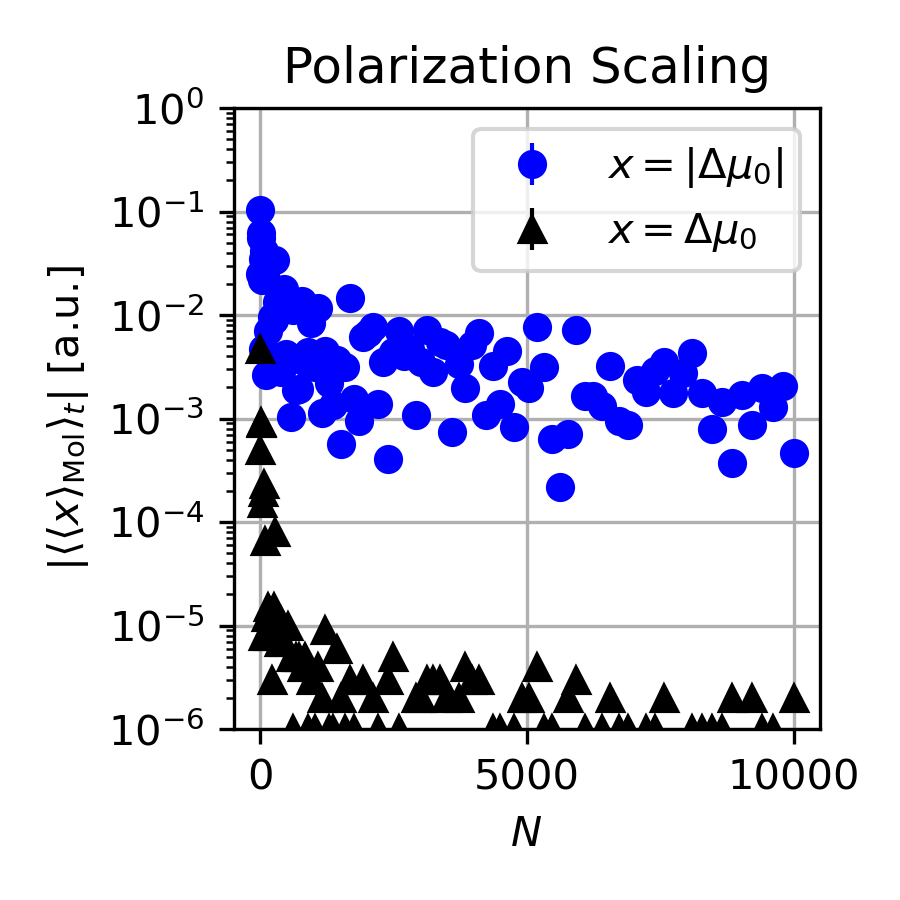}
         %\caption{}
         %\label{fig:scaling}
     %\end{subfigure}
     %\hfill
     %\begin{subfigure}[b]{0.45\textwidth}
      %   \centering
      %   \includegraphics[width=\textwidth]{dr_e_distribution_N1024_l0004_n1200nice_.png}
         %\caption{   }
     %\end{subfigure}
     \caption{Local polarization features of Shin-Metiu molecules under collective VSC.  Finite cavity-induced molecular polarizations (blue) that emerge due to the self-consistency cycles of the cavity Hartree equations (reproduced from Ref.~\citenum{sidler_unraveling_2024}). The collective Rabi splitting was kept fixed, when increasing the ensemble size $N$.  In contrast, the macroscopic polarization (black) quickly drops to zero, within numerical errors. Notice that the numerical results of the Shin-Metiu molecules were determined at the onset of a numerical instability (phase transition), which effectively prevents to reach considerably stronger collective coupling strength with the chosen setup. %\textbf{Right:} Probability density of the cavity-induced polarizations calculated for $N=1024$ molecules of the above setup. The system was equilibrated for $600$ time steps and then polarization data was considered from another $600$ time steps. Two striking features appear: First, the local polarizations are symmetrically distributed. Second, the probability distribution seems almost discontinuous, even though the electronic-structure data is time-averaged over many vibrational cycles. This should smoothen the data, unless some frustration mechanism counteracts.
     }
     \label{fig:distribution}
\end{figure}

While the previous analytical results already show local effects under VSC for a externally driven system, we are interested specifically in the thermal equilibrium properties of molecular ensembles under VSC. To that end, one has to propagate the classical equations of motion from Eq.~\eqref{eq:class_ham} in contact with a thermal bath.~\cite{sidler_perspective_2022} To connect to the macroscopic experimental setup, one usually considers a fixed collective Rabi splitting (see, e.g., Fig.~\ref{fig:splitting}) and then slowly increases the number of molecules. That is, the macroscopic observable in absorption or transmission is the appearance of two peaks where there was one (highly degenerate) vibrational peak of the free-space ensemble. The distance between those two peaks is called the Rabi splitting and it is a measure of \textit{collective} coupling strength $\lambda_{\rm coll}$ of the macroscopic ensemble. Having fixed the observed Rabi splitting, we perform a thermodynamic limiting procedure to see how different ensemble property behaves, and hence use $\lambda = \lambda_{\rm coll}/\sqrt{N}$ with $N\rightarrow \infty$. As indicated in Sec.~\ref{sec:pauli}, this limiting procedure should be taken with a grain of salt, since a cavity has a finite mode volume and the effective coupling strength is always non-zero. Thus, if the property does not converge for a given mode volume $\mathcal{V}$, then the presented level of description is most likely insufficient~\cite{svendsen_ab_2024}. 
\\
%\\
Turning back to the harmonic model, the above procedure converges, however, to a rather boring result: The local effects vanish in the thermodynamic limit at thermal equilibrium~\cite{horak_analytic_2024}. Of course, the setting is oversimplified and it is known that harmonic models often miss important physical aspects. So we consider a slightly more realistic example where the (effective) electron and nuclei do not interact harmonically.
%Traditionally such a classical molecular dynamics setup would be considered overall in (canonical) thermal equilibrium.~\cite{hutter_carparrinello_2012} However, things can become more intricate in a cavity~\cite{sidler_perspective_2022}, e.g., when forming a glassy phase as will be discussed in Sec.~\ref{sec:resonance}. For each propagation time step the cavity Hartree Eqs.~\eqref{eq:cavhartree} have to be solved self-consistently for all the electrons. This is numerically very expensive and demanding. Already minimizing the  electronic-structure problem of a single realistic molecule at a single time-step in principle poses a formidable challenge.~\cite{schuch_computational_2009} 
For this reason, an ensemble of an-harmonic Shin-Metiu model molecules~\cite{shin_nonadiabatic_1995} was used in Ref.~\citenum{sidler_unraveling_2024}. The Shin-Metiu molecule is a paradigmatic and common model to study chemical reactions and conical intersections in- and outside of cavities~\cite{li_cavity_2021,hu_quasi-diabatic_2022,fischer_beyond_2023,mandal_theoretical_2023,zhu_making_2024}. 
%Each Shin-Metiu molecule consists of only one effective electron and nucleus and thus allows a numerically efficient solution of the free-space matter problem. 
This computationally simple model permits the efficient exploration of classical cavity-Born-Oppenheimer molecular dynamics at finite temperature up to several 1000 molecules.~\cite{sidler_unraveling_2024} %Most prominently, the numerical solution allows to relax the two-level restriction of the  electronic-structure problem prevalent in the SK mapping.
%Figs. \ref{fig:splitting} and \ref{fig:scaling} are re-printed from Ref. \citenum{sidler2023unraveling}. 
In Fig.~\ref{fig:splitting} we see the corresponding collective Rabi splitting (dashed blue line) for $N=900$. The lower peak is called the collective lower polariton and the upper peak the collective upper polariton, respectively. These peaks correspond to hybrid cavity-matter excitations, from which the field of polaritonic chemistry inherits its name.~\cite{ebbesen_hybrid_2016,garcia-vidal_manipulating_2021,ebbesen_introduction_2023} The Rabi splitting is asymmetric, since the self-consistent solution leads to a change in refractive index (red-shift), as also discussed above for the harmonic model (see Eq.~\eqref{eq:redshift}). 
Apart from these macroscopic properties, Fig.~\ref{fig:splitting} also shows the averaged hypothetical absorption spectrum of the average single molecule in the ensemble (solid blue line). Interestingly we find besides the dark states\cite{sidler_polaritonic_2021,sidler_unraveling_2024,groenhof_tracking_2019,davidsson_atom_2020,davidsson_role_2023, ulusoy_dynamics_2020,du_catalysis_2022,gonzalez-ballestero_uncoupled_2016,delpo_polariton_2020,campos-gonzalez-angulo_generalization_2022,borges_extending_2024,borges_role_2025} (vibrational states that decouple from the photon field) that also the local lower polariton is populated. As specific quantity that we want to consider in the above thermodynamic limiting procedure we consider the cavity-induced polarization differences 
\begin{align}
\Delta\mu_0 = \langle \hat{\mu}_i\rangle_{0,\lambda=0}-\langle\hat{\mu}_i\rangle_{0,\lambda}.
\end{align}
for increasing number of molecules. Here the electronic polarization operator of the $i$-th Shin-Metiu molecule is given by $\hat{\mu}_i=-Z_e \hat{r}_i$. Indeed, Fig.~\ref{fig:distribution}  is the first theoretical evidence of a cavity-induced local equilibrium polarization mechanism under collective VSC.~\cite{sidler_unraveling_2024} In more detail, the time and ensemble averaged polarizations of the single molecules approach a finite value in the thermodynamic limit (blue). In contrast, the macroscopic polarization (black) quickly drops to a vanishingly small value. Notice, these findings hold qualitatively for an ensemble of aligned as well as randomly oriented molecules.~\cite{sidler_unraveling_2024} 
The polarization pattern revealed by Fig.~\ref{fig:distribution} can be summarized as
\begin{eqnarray}
    E_{\rm cH}[\Delta\mu_0]&=&0,\label{eq:ch_mean}\\
    {\rm Var}_{\rm cH}[\Delta\mu_0]&\neq& 0,\label{eq:ch_var}
\end{eqnarray}
which vaguely resembles a spin glass phase, for which, simply speaking, one observes zero overall magnetization, but ordering of the local spins (magnetization).~\cite{edwards_theory_1975,sherrington_solvable_1975}
For this reason, the above local polarization pattern with zero net polarization was termed a \textit{polarization glass} in our previous work in Ref.~\citenum{sidler_unraveling_2024}.
%Naturally, the question arises why do these small single-molecule polarizations not add up constructively to generate a macroscopic polarization? 
%so far the absence of a cavity-induced macroscopic polarization has not been explained sufficiently from the numerical results. It can either be a consequence of the randomly oriented/aligned molecules with finite local polarization, or the single molecular polarization is elongated and squeezed at the same time, such that they balance each-other. %The later case, cannot be represented with two local basis functions as assumed in the SK model, where by construction any local polarization favours a direction. 
 
%
%Furthermore, we notice that the probability density appears relatively rough (almost discontinuous), even though it is an averaged quantity over many vibrational cycles (nuclear time scales). The emergence of such almost discontinuous probability distribution, as well as the very noisy data for the cavity-induced local polarizations (even on a logarithmic scale), suggest the presence of a \textit{frustration} mechanism, i.e., the local polarization patterns are long-lived, despite being constantly perturbed by the vibrating nuclei. Moreover, the frustrated behavior suggests that the time- and ensemble-averaged data is strongly correlated with the randomly chosen initial state of the system. This is what one would expect in a spin glass as well.~\cite{parisi_nobel_2023} 

Eventually, we would like to highlight that solving the cavity Hartree Eq. (\ref{eq:cavhartree}) for the Shin-Metiu molecules is only possible up to a certain collective Rabi splitting. When increasing the collective coupling strength beyond this value, the self-consistency cycles of the cavity Hartree equations do not converge anymore.~\cite{sidler_unraveling_2024} This suggests the emergence of a cavity-induced \textit{degeneracy/phase transition} at a certain collective coupling strength. Again, the occurrence of a highly degenerate ground state that depends on the strength of an all-to-all interaction term closely resembles the behavior of a spin glass. In the following we want to further explore the resulting implications including electron correlation effects.

%But before, we discuss some experimental evidence for the proposed phase transition.
%Exploring the theoretical connection between the polarization glass and the spin glass picture and thus unifying the polaritonic chemistry with the established field of spin glasses will be the major target of this review article.  


\section{Mapping cavity-mediated molecular electron correlations to a spin glass~\label{sec:SSKmodelemerges}}
%\revDS{Connect from HF (exact?) in large N limit to Single excitations. In particular it needs to become clear that doubles, triple excitations are irrelevant. Either large N argument or restricting to quasi degenerate states or both could help. However, ideally we do not restrict ourselves here to degeneracies... }

We have seen that local modifications can survive in the large-$N$ limit even at equilibrium if we go beyond the harmonic approximation. Thus under collective VSC, electronic fluctuations are enhanced when compared to the free-space case. The collective SCF convergence issue suggests that many electronic configurations have roughly the same ensemble energy. Thus one runs into an energetically very dense \textit{intermolecular} spectral region and hence faces huge (at least numerical) degeneracies. That such a situation appears is to be expected, because on the macroscopic ensemble level, entropic/statistical contributions become important. That is, we need to weigh the energy with its number (density) of states and perform free-energy considerations. In order to do so, we need to get information on the electronic density of states, i.e., we need to keep track of how many excited ensemble states are near to the (potentially highly degenerate) ground state.
\\
\\
In quantum chemistry, there is a simple way to get this information. One can perform a configuration-interaction (CI) singles calculations on top of Hartree-Fock Slater determinants~\cite{szabo_modern_2012}, i.e., we include electron exchange and correlations. To do so, we leave the dilute-gas assumption and consider the full cHF problem of Eq.~\eqref{eq:cHF}. That is, the  electronic-structures of the individual molecules can overlap and we consider the full cHF ensemble wave function. We remind the reader that on the \textit{intermolecular} scale, the cHF wave function is supposed to become asymptotically exact, and hence we can neglect CI doubles and triples for the ensemble problem. Before we continue, with CI singles, we briefly would like to mention that over the past years different post-Hartree-Fock methods were proposed to account for cavity-modified correlation effects. For example, coupled-cluster\cite{mordovina_polaritonic_2020,haugland_coupled_2020,deprince_cavity-modulated_2021}, Moeller-Plesset perturbation theory\cite{bauer_perturbation_2023,cui_variational_2024,matousek_polaritonic_2024,el_moutaoukal_strong_2025}, configuration interaction,\cite{mctague_non-hermitian_2022} multi-configurational\cite{vu_cavity_2024,alessandro_complete_2025}, DFT-based \cite{ruggenthaler_time-dependent_2011,ruggenthaler_quantum-electrodynamical_2014,pellegrini_optimized_2015,flick_ab_2018,tokatly_time-dependent_2013,lu_electron-photon_2024,tasci_photon_2025} and other correlated methods\cite{weight_diffusion_2024,mallory_reduced-density-matrix-based_2022,matousek_polaritonic_2024} are known. However, typically, ab-initio methods that are designed to accurately capture electron correlations are naturally restricted to only a few molecules. A notable generalization to the collective regime can be found in Ref.\citenum{castagnola_realistic_2025} for a coupled cluster-model, which seems to accurately describe local (e.g. intra-molecular) correlation energies under collective electronic strong coupling in a dilute gas. Alternatively, a radiation-reaction embedding approach was proposed in Ref. \cite{schafer_polaritonic_2022} or machine learning of forces\cite{schafer_machine_2024}. However, the later two approaches sacrifice the collectively-induced\cite{sidler_unraveling_2024} feed-back effects and thus do not account for collectively induced degeneracies and its consequences on the inter-molecular electron correlations. Instead, our subsequent theoretical framework of a polarization/spin glass  is fundamental for  describing polaritonic chemistry. The self-consistency is an essential ingredient for the understanding on the local vs. collective interplay of realistic molecular ensembles under collective VSC.
%Previously, we have seen that intermolecular cHF features are analytically interpretable in the dilute gas limit, where absence of electronic overlap allows for a Hartree-product Ansatz, made of $N$ molecular wave-functions. In a next step, we increase the density of our molecular ensemble. Consequently, Hartree exchange as well as intermolecular electron correlations will start to play a relevant role.
%1st) assume quasi dilute gas limit from now on. Probably move discussion what it means from bottom to here.
%\begin{eqnarray}
%\hat{H}^e=\hat{h}_m(\boldsymbol{R})+\hat{h}_c(\boldsymbol{R},q_\beta)+ \frac{1}{2}\sum_i^{N}\sum_{j\neq i}^{N}\boldsymbol{\lambda}\cdot\boldsymbol{r}_i\ \boldsymbol{\lambda}\cdot\boldsymbol{r}_j
%\end{eqnarray}

For a given cHF solution $\ket{\Psi_0}$ (reference ensemble determinant), the the lowest lying excited states are found by~\cite{szabo_modern_2012}
\begin{eqnarray}
    \ket{\Phi_{\boldsymbol{s}}}= \sum_{c}^N\sum_{t}^\infty s_c^t \ket{\Psi_c^t}, 
\end{eqnarray}
where $\boldsymbol{s}$ are the occupations of the different Slater determinants. Here $\ket{\Psi_c^t}$ is a singly excited Slater determinant, where the occupied spin-orbital $c$ is excited to the $t$-th unoccupied orbital. The usual orthogonality conditions between the excited Slater-determinants hold $\braket{\Psi_c^t}{\Psi_d^u}=\delta_{c,d}\delta_{t,u}$ and like-wise with respect to $\ket{\Psi_0}$.~\cite{szabo_modern_2012}
To simplify our discussion further, we restrict our considerations to closed-shell cHF in the following, i.e., the spin orbital $\phi_i(\boldsymbol{\tau})$ representation of the electron integrals turn into a spatial orbital representation $\chi_i(\boldsymbol{r})$. In particular, every orbital is assumed to be doubly occupied by electrons. To simplify our notation, the spatial DSE and Coulomb two-electron integrals are given as follows:
\begin{eqnarray}
    (ud|ct)_{\rm DSE}&=&\langle\chi_u|\boldsymbol{\lambda}\cdot\hat{\boldsymbol{r}}|\chi_d\rangle \langle\chi_c|\boldsymbol{\lambda}\cdot\hat{\boldsymbol{r}}|\chi_t\rangle,\\
     (ud|ct)_{\rm C}&=&\int d\boldsymbol{r}_1 d\boldsymbol{r}_2\chi_u^*(\boldsymbol{r}_1 )\chi_d(\boldsymbol{r}_1 )|\hat{\boldsymbol{r}}_1-\hat{\boldsymbol{r}}_2|^{-1} \chi_c^*(\boldsymbol{r}_2 )\chi_t(\boldsymbol{r}_2 ),
\end{eqnarray}
with $ (ud|ct)= (ud|ct)_{\rm C}+(ud|ct)_{\rm DSE}$.
We note that the standard Brillouin theorem $\langle \Psi_0|\hat{H}^e|\Psi_c^t\rangle=0$, as well as the Slater-Condon rules remain valid for the DSE two-electron integrals. Consequently, the transition matrix element between singlet symmetry-adapted configurations from real orbitals reduce to the following compact form~\cite{szabo_modern_2012} 
\begin{eqnarray}
    \langle\Psi_d^u|\hat{H}^e|\Psi_c^t\rangle=(E_0+\epsilon_t-\epsilon_c)\delta_{d,c}\delta_{u,t}+2(ud|ct)-(ut|cd).\label{eq:signletransition}
\end{eqnarray}
The resulting CI-singles energies are
\begin{align}\label{eq:total_energy}
    \mathcal{E}_{\boldsymbol{s}}&=\langle\Phi_{\boldsymbol{s}}|\hat{H}^e|\Phi_{\boldsymbol{s}}\rangle\\
    &=\frac{1}{\sum_{e v}(s_e^v)^2} \sum_{c}^N\sum_{d}^N\sum_{t}^\infty\sum_{u}^\infty \bigg( s_c^t s_d^u \big[(E_0+\epsilon_t-\epsilon_c)\delta_{d,c}\delta_{u,t}+2(ud|ct)-(ut|cd)\big]\bigg).\nonumber 
\end{align}
By ordering all the CI-singles energies we can then explore the lowest lying excitations and also potential degeneracies.
\\
\\
To again make things a little simpler and focus on the essentials, i.e., the long-range interactions induced by the cavity, we make a \textbf{quasi-dilute gas} assumption.
That is, we will partition our orbitals $c,d,t,u$ into a set $S_{\rm inter}$, for which we have
\begin{align}
2(ud|ct)_{\rm C}-(ut|cd)_{\rm C} \rightarrow 0 \quad {\rm yet} \quad 2(ud|ct)_{\rm DSE}-(ut|cd)_{\rm DSE} \neq 0.
\end{align}
At the same time we assume that the rest of the orbitals are in $S_{\rm intra}$, for which 
\begin{align}
2(ud|ct)_{\rm C}-(ut|cd)_{\rm C} \neq 0  \quad {\rm yet} \quad 2(ud|ct)_{\rm DSE}-(ut|cd)_{\rm DSE} \rightarrow 0.
\end{align}
%with the following properties, i.e,  
%\begin{eqnarray}
% c,d,s,u\ \in S_{\rm quasi-dilute}=\{ (ud|ct)_{\rm C}\rightarrow 0  \wedge   
%(ud|ct)_{\rm DSE}\neq 0 \}.
%\end{eqnarray}
In other words, we assume a partitioning of our set of orbitals into \textit{intramolecular} orbitals in $S_{\rm intra}$, which are the focus of standard (single-molecule) quantum chemistry, and into \textit{intermolecular} orbitals in $S_{\rm inter}$, which are de-localized orbitals. For the latter, the long-range properties of the transverse DSE interaction start to dominate over the short-ranged (longitudinal) Coulomb interaction. For localized molecular orbitals, which determine the \textit{intramolecular} structure, the Coulomb two-electron integrals will be decisive and thus cannot be discarded. 
Thus Eq.~\eqref{eq:total_energy} can be separated as 
\begin{eqnarray}
    \mathcal{E}_{\boldsymbol{s}}&=&E_0+E_{\boldsymbol{s}}^{\rm C}+E^{\rm DSE}_{\boldsymbol{s}} \nonumber \\
    &\approx& E_0+E^{\rm C, intra}_{\boldsymbol{s}}+E^{\rm C, inter}_{\boldsymbol{s}}+E^{\rm DSE, inter}_{\boldsymbol{s}} \nonumber \\
    &\overset{\rm quasi\ dilute}{\approx}&E_0+E^{\rm C, intra}_{\boldsymbol{s}}+E^{\rm DSE, inter}_{\boldsymbol{s}},\label{eq:quasidilutcorr}
\end{eqnarray}
where the first approximation is a consequence of the typically small contributions of the DSE interaction for localized orbitals, whereas the second approximation is only reasonable, if the quasi-dilute gas approximation applies for all molecules within our ensemble of interest. 

Using the quasi-dilute gas picture in Eq.~\eqref{eq:quasidilutcorr}, we can consider the DSE correlation energy $E^{\rm DSE, inter}_{\boldsymbol{s}}=\big(\sum_{c,d,t,u\ \in S_{\rm inter}} s_c^t s_d^u \big[(\epsilon_t-\epsilon_c)\delta_{d,c}\delta_{u,t}+2(ud|ct)-(ut|cd)\big]\big)/\big(\sum_{e v}(s_e^v)^2\big)$ independent from the local intramolecular correlation energy described by $E^{\rm C, intra}_{\boldsymbol{s}}=\big(\sum_{c,d,t,u\ \in S_{\rm intra}} s_c^t s_d^u \big[(\epsilon_t-\epsilon_c)\delta_{d,c}\delta_{u,t}+2(ud|ct)-(ut|cd)\big]\big)/\big(\sum_{e v}(s_e^v)^2\big)$. That is, the intra- and the intermolecular energies decouple and we can exclusively focus on the intramolecular energy contributions. In a next step we note that 
\begin{align}
    \sum_{c}^N\sum_{d}^N\sum_{t}^\infty\sum_{u}^\infty s_c^t s_d^u (ud|ct)_{\rm DSE} = %\sum_{u,d} \bigg(s_d^u \langle\chi_u|\boldsymbol{\lambda}\cdot\hat{\boldsymbol{r}}|\chi_d\rangle \bigg) 
    \bigg(\sum_{c,t} s_c^t \langle\chi_c|\boldsymbol{\lambda}\cdot\hat{\boldsymbol{r}}|\chi_t\rangle \bigg)^2\geq0.\label{eq:partitioning}
\end{align}
%This bi-partite partitioning holds likewise for the restriction to the subset $S_{\rm quasi-dilute}$.
%The first term on the right of Eq.~\eqref{eq:partitioning} is neglected in the following. The reason is that for our setup the coefficients $c_c^t$ must be real valued. Hence this term is positive semi-definite, i.e., it can only increase the DSE correlation energy. Since we want to minimize the total energy, setting it to zero, comes at the cost of imposing only one additional global constraint on the coefficients $c_c^t$ that mediate DSE long-range interactions. Using that the second term in Eq.~\eqref{eq:partitioning} becomes diagonal with respect to our representation in terms of singly-excited Slater determinants. The resulting DSE correlation energy problem can thus be written as follows:
That is, the correlations have a similar bi-partite Hartree term as the cHF equation for $\ket{\Psi_0}$. This allows to restructure the DSE correlation energy as
\begin{align}
    E^{\rm DSE, intra}_{\boldsymbol{s}}=&
    \frac{1}{\sum_{e v}(s_e^v)^2}\bigg[\sum_{c,t}
(s_c^t)^2 \underset{= -\mathcal{K}_{cc}^{tt}}{\underbrace{\big(\epsilon_t-\epsilon_c-(tt|cc)_{\rm DSE}\big)}} -\sum_{c,d}\sum_{t,u}s_c^t s_d^u \underset{\mathcal{K}_{cd}^{tu}}{\underbrace{(ut|cd)_{\rm DSE} (1-\delta_{cd}\delta_{tu})}} \nonumber \\
&  + \bigg(\sum_{c,t} s_c^t\langle\chi_c|\boldsymbol{\lambda}\cdot\hat{\boldsymbol{r}}|\chi_t\rangle \bigg)^2%\sum_{c,t} \sum_{u,d} s_c^t s_d^u  \langle\chi_u|\boldsymbol{\lambda}\cdot\hat{\boldsymbol{r}}|\chi_d\rangle \langle\chi_c|\boldsymbol{\lambda}\cdot\hat{\boldsymbol{r}}|\chi_t\rangle\bigg]
\label{eq:Kdef}\\
&=-\sum_{c,d}\sum_{t,u}s_d^u s_c^t\frac{\mathcal{K}_{cd}^{tu}}{\sum_{e v}(s_e^v)^2}+ \frac{2}{\sum_{e v}(s_e^v)^2}\bigg(\sum_{c,t} s_c^t\langle\chi_c|\boldsymbol{\lambda}\cdot\hat{\boldsymbol{r}}|\chi_t\rangle \bigg)^2 \label{eq:KKform}
\end{align}
Since the second term of Eq.~\eqref{eq:KKform} is positive and thus its minimal value is zero, when searching for the lowest correlation energies, this term can be incorporated as a side condition that fixes one occupation number $s_d^u$. That is, we can perform a minimzation by enforcing $\sum_{c,t} s_c^t \langle\chi_c|\boldsymbol{\lambda}\cdot\hat{\boldsymbol{r}}|\chi_t\rangle = 0$. We will thus discard this energy contribution and only take it into account by restricting the space of allowed states to vary over. In a next step we re-label the orbital occupations by
\begin{align}
s_{i} = \frac{s_{c}^{t}}{\sqrt{\sum_{e,v} (s_{e}^{v})^2}},
\end{align}
and the overlaps by
\begin{align}
\mathcal{J}_{ij} = \frac{\mathcal{K}_{cd}^{tu}}{\sum_{e v}(s_e^v)^2},
\end{align}
such that we end up with
\begin{align}
E^{\rm DSE, intra}_{\boldsymbol{s}}=-\sum_{i,j}s_i s_j \mathcal{J}_{ij}.\label{eq:spinglass}
\end{align}
For sufficiently large molecular ensembles within the quasi-dilute gas regime, the resulting $\mathcal{J}_{ij}$ can be regarded as a random variable. Its distribution depends on the respective orbital excitation energy $\epsilon_t-\epsilon_c\mapsto \Delta\epsilon_i$, as well as on the orientation (polarization) of the the excitation with respect to the polarization of the relevant cavity modes, (diagonal and off-diagonal contributions from $(ut|cd)_{\rm DSE}$-terms, see Eq.~\eqref{eq:Kdef}). As a consequence, Eq.~\eqref{eq:spinglass} is similar to a \textbf{spin glass} Hamiltonian.~\cite{mezard_spin_1987,kosterlitz_spherical_1976} The probability distribution of $\mathcal{J}_{ij}$ will strongly depend on the molecular properties of the collectively coupled ensemble. In more detail, to expect non-negligible DSE correlation effects (i.e., $\exists \; i\neq j,\ {\rm with}\ |\mathcal{J}_{ij}|> 0$), very de-localized electronic orbitals (excitations) are required. However, most of this de-localized excitations will also possess a high orbital excitation energy $\Delta \epsilon_i$ and can thus be safely discarded from the considerations in Eq.~\eqref{eq:spinglass} when exploring the lowest electronic energy landscape. Consequently, to expect significant \textit{intermolecular} DSE correlation effects, we additionally require a highly (almost) degenerate HF ground state (i.e.  $\exists i,\ {\rm with}\ \langle\mathcal{J}_{ii}\rangle\approx 0$). For sufficiently strong collective coupling, we have already seen that such a highly degenerate electronic ground state (polarization glass) exists in the dilute gas limit (see Sec. \ref{subsubsec:polarizationglass}). Increasing slightly the \textit{intermolecular} densities, i.e., going to the quasi-dilute limit, we expect that at least for specific chemical setups we find a non-vanishing \textit{intermolecular} DSE correlation energy of the following form
\begin{eqnarray}
    E_{\rm corr}^{\rm DSE} =-\sum_{i<j}^{N_J}s_i s_j J_{ij},\ \sum_{i }s_i^2=1,\ \langle J_{ii}\rangle=0,\label{eq:spinglass_gen}
\end{eqnarray}
where $N_J$ is the number of the relevant (almost zero energy when compared to the relevant scale) orbital excitations in the ensemble. 
We expect the distribution of the random variables $J_{ij}$ to be a heavily-tailed distribution (e.g. Cauchy-like), since most excitations will contribute only a little, but there might be a few very de-localized excitations, which contribute significantly. Understanding the spin glass properties of Eq.~\eqref{eq:spinglass_gen} for such distributions will require non-trivial and computationally expensive simulation. Finally, we note that a cavity-mediated spin glass phase is not an entirely new concept. However, previous theoretical concepts as well as experimental realizations rely on the complex restructuring of the photon modes to reach the desired random interactions.\cite{gopalakrishnan_frustration_2011, gopalakrishnan_exploring_2012,guo_sign-changing_2019,guo_emergent_2019,marsh_enhancing_2021,marsh_entanglement_2024,kroeze_replica_2023}  Instead, the here proposed concept utilizes the complex  electronic-structure of molecules to create a cavity-mediated spin glass. In more detail, the origin of randomness emerges from the interplay between breaking the isotropy of space and orienting the molecules randomly with respect to the distinguished polarization axis.

%%%%%%%%%%%%%%%%%%%%%%%%%%%%%%%%%%%%%%%%%%%% CONtinue original manuscript after new mapping is iserted

\section{The physics of a spin glass\label{sec:spin_glass}}

In the previous sections we have established a formal connection between the physics of spin glasses and the cavity-induced inter-molecular electron correlations under VSC. Thanks to the theoretical similarities between both problems, we are confident that the polaritonic-chemistry community can learn from established knowledge of the spin glass community. In the following, we briefly introduce some key-concepts of spin glasses using the  Sherrington-Kirkpatrick (SK) model. It provides a paradigmatic model of a spin glass with long-range (all-to-all) interaction, for which exact results have been determined after decades of intensive research that ultimately were awarded with the Nobel prize in physics (2021).~\cite{parisi_nobel_2023} 
We note that many models of spin glasses are based on (random) short-ranged interactions between spins on a crystal lattice, e.g., the Edwards-Anderson model~\cite{edwards_theory_1975}. Furthermore, the classical spin variable are usually discrete.  However, in a cavity we have a continuous long-range (all-to-all) electron correlation interaction. %and a differently scaling coupling to the displacement and nuclear coordinates. 
The spin glass properties of the SK model are considered generic, since it serves as a limiting case~\cite{parisi_spin_2006} for the Bethe lattice model,~\cite{bethe_statistical_1935} the long-range Edwards-Anderson model and the infinite dimensional Edwards-Anderson model.~\cite{edwards_theory_1975}. Those key concepts, for example, involve \textbf{spin glass phase transition, frustration, replica symmetry breaking, (off)-equilibrium fluctuations and aging effects}.  
Eventually, after having introduced above concepts for the SK model, we will have a closer look at the spherical 2-spin glass model,\cite{kosterlitz_spherical_1976} that provides (to our knowledge) the closest known spin glass model to describe the cavity-mediated electron correlations derived in Eq.~\eqref{eq:spinglass_gen}.

\begin{figure}
     %\begin{subfigure}[b]{0.45\textwidth}
         \centering
         \includegraphics[width=0.45\textwidth]{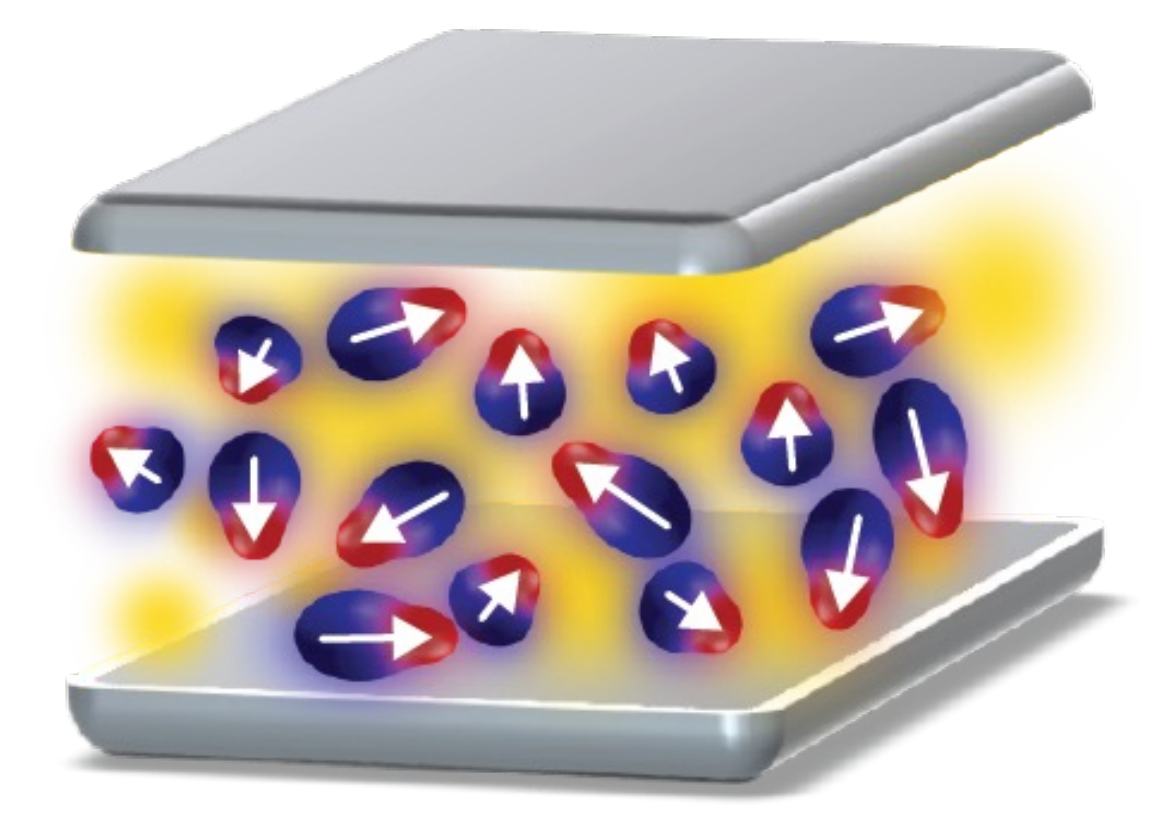}
         \caption{Pictorial representation of a  cavity-mediated  molecular polarization/spin glass under by vibrational strong coupling. White arrows indicate the locally modified (slighly polarized)  electronic-structure. Figure taken from Ref.\citenum{sidler_unraveling_2024}.   }
         \label{fig:spin_vs_polarization_glass}
     %\end{subfigure}
\end{figure}

\subsection{The Sherrington-Kirkpatrick (SK) model}

The Hamiltonian of the SK model of a spin glass is given by
\begin{eqnarray}
   % H^{\rm e}_{\boldsymbol{R},q_\beta}(\boldsymbol{\sigma})\sim
   H_{SK}(\boldsymbol{\sigma})=- \sum_{j<i}^{N_J} J_{ij}\sigma_i\sigma_j,\ ,\ J\sim\mathcal{N}(J_0/N_J,\tilde{J}^2/N_J) %+h\sum_i \sigma_i 
    \label{eq:sk_model}
\end{eqnarray}
and the couplings $J_{ij}$ between the spins are independent random variables that are normally distributed. The spins take discrete values $\sigma_i \in\{\pm1\}$. The presence of an additional external magnetization field $h$, acting on the spins as $\sum_i^{N_J} h \sigma_i$ to generate a finite magnetization $m$, can be recast into a finite mean-value $J_0=h/m$ of the random interactions $J_{ij}$.~\cite{almeida_stability_1978} 
To understand the basic physical properties of the SK model at finite temperature $T$, we follow Refs.~\citenum{parisi_nobel_2023,parisi_spin_2006} and continue with some definitions. The local ("single molecule") magnetization (in the case of polaritonics this would be polarization) at temperature $T$ is given by, 
\begin{eqnarray}\label{eq:localmagentization}
    m(i)_\alpha= \langle \sigma_i\rangle_{T,\alpha},
\end{eqnarray}
%\revDS{Ruggi: do you have a good suggestion for the different averages. Here it would be a thermal state, but not averaged over alpha and J}
where $\alpha$ denotes a possible quasi-thermal equilibrium state of the SK model, i.e., a local minimum in the phase space for a given choice of $J$, at temperature $T$. In more detail, the  magnetizations $ m(i)_\alpha$ correspond to the $\alpha$-th solution of the mean-field equation,~\cite{thouless_solution_1977,parisi_spin_2006}
\begin{eqnarray}
    m(i)=\tanh \big (\frac{\sum_j J_{ij}m(j)}{k_B T}\big),\label{eq:meanfield_m}
\end{eqnarray}
which becomes an exact description for the SK model in the thermodynamic limit.~\cite{sherrington_solvable_1975, parisi_infinite_1979, parisi_spin_2006} 
The determination of the exponentially large number of solutions is a computationally very demanding task. 
Eventually, the thermal equilibrium ("single molecular") magnetization, averaged over all $\alpha$ and all possible choices of J, is defined as
\begin{eqnarray}
    m=\langle\langle m(i)_\alpha\rangle_\alpha\rangle_J.\label{eq:magnetization}
\end{eqnarray}
We note that averaging over all $\alpha$ and $J$ makes the result site-independent, i.e., the left-hand side of Eq.~\eqref{eq:magnetization} is independent of $i$. The numerous fundamental physical properties of the SK model and its implications the understanding of spin glasses in general will be discussed in the following with its implications on polaritonic chemistry (see Sec. \ref{sec:consequences}).

\subsubsection{Spin glass phase\label{sec:instability}}

\begin{figure}[h]
\includegraphics[width=0.5\textwidth]{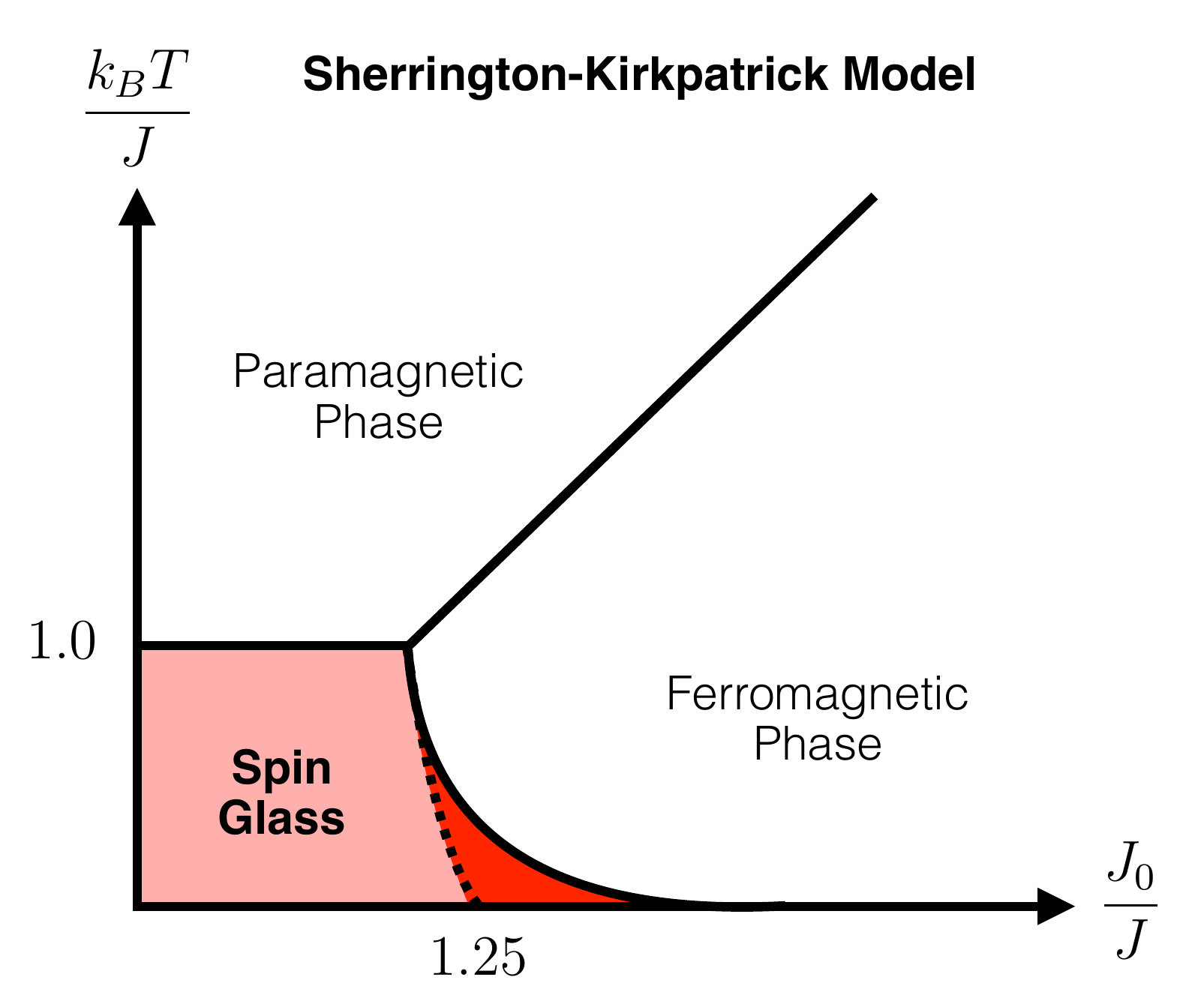}
\caption{ Phase diagram of the SK model, reproduced from Ref. \citenum{almeida_stability_1978}. The doted line corresponds to the original (erroneous) spin glass phase proposed by Sherrington and Kirkpatrick, which did not consider that spontaneous replica symmetry breaking can occur. Replica symmetry breaking extends the (unstable) spin glass regime at low temperature to much larger $J_0/J\gg 0$ values (dark red). Notably, in the ground state ($T\rightarrow0$), the instability occurs for arbitrarily large (but finite) $J_0$-values, provided that $J>0$.  %Under the VSC mapping typically  $J_0\propto h_m\gg 0$, which indicates that the electronic temperature $k_B T\ll J$ must be sufficiently low (blue) such that  the bare matter solution becomes unstable for moderate collective coupling strengths $J$. If this is the case, the cavity induced polarization glass phase is entered, which triggers not only local electronic changes (represented by local spin-flips), but also thermodynamic modifications that are generic for spin glasses (see subsequent discussions).  }
}
\label{fig:phasediagram}
\end{figure}

To determine the phase diagram of the SK model, we introduce the Edwards-Anderson order parameter or self-overlap as defined in Ref. \citenum{edwards_theory_1975} 
\begin{eqnarray}
     q_{EA}=\frac{\sum_i m(i)_\alpha m(i)_\alpha}{N}={\rm const}. \ \forall \alpha,J,
 \end{eqnarray} 
which can be shown to neither depend on the state $\alpha$ nor the specific realization of J.\cite{parisi_spin_2006,parisi_nobel_2023} 
Based on the magnetization $m$ and the magnetic order parameter  $q_{EA}$, Sherrington and Kirkpatrick determined an analytical phase diagram in the thermodynamic limit $N\rightarrow\infty$. Their computations suggested the emergence of three different phases:  a ferromagnetic ($q_{EA}\neq 0$, $m\neq 0$), a spin glass  ($q_{EA}\neq 0$, $m = 0$) and a paramagnetic one ($q_{EA}=0$, $m= 0$).   However, Sherrington and Kirkpatrick already noticed that their solution seemed questionable at low temperatures, since it gave rise to a negative entropy, which is nonphysical.\cite{sherrington_solvable_1975} Indeed, Almeida and Thouless showed that the original solution of the SK model becomes unstable in the thermodynamic limit at sufficiently low temperature, which leads to the corrected phase diagram of the SK model displayed in Fig.~\ref{fig:phasediagram}.~\cite{almeida_stability_1978}  Almeida and Thouless could determine an explicit stability criterion at low temperature for the SK model~\cite{almeida_stability_1978}
\begin{eqnarray}
    k_B T > \frac{4}{3\sqrt{2 \pi}}J e^{-\frac{J_0^2}{2 J^2}}.\label{eq:instability}
\end{eqnarray}
The striking feature of this result is that in the ground state ($T\rightarrow 0$), the solution of the SK model becomes unstable, and thus enters the spin glass phase, even if $\infty>J_0\gg J$, i.e., even if the system is exposed to very strong external magnetization fields. 
\\
\\
For a polaritonic setting this could imply that even a (very) small intermolecular DSE correlation interaction can potentially introduce a spin glass-type phase transition that fundamentally alters microscopic and macroscopic properties of a polaritonic ensemble at low-enough temperatures. The SK analogy also highlights that the temperature for the electronic subsystem might become important when treating the dressed  electronic-structure under VSC. Even though we assumed bare molecules, for which non-adiabatic coupling effects to the \textit{intramolecular} excited  electronic-structures are irrelevant in free space. 
\\
\\
The origin of the Almeida and Thouless instability can be attributed to spontaneous replica symmetry breaking that we briefly discuss next.\cite{parisi_nobel_2023} Notice further, that a spin glass is conceptually distinct from the Anderson localization mechanism, which has been discussed in the context of polaritonic transport properties.~\cite{schachenmayer_cavity-enhanced_2015,hagenmuller_cavity-enhanced_2017}

\subsubsection{Free energy and spontaneous replica symmetry breaking \label{sec:re_breaking} }

\begin{figure}
         \begin{subfigure}[b]{0.45\textwidth}
         \centering
         \includegraphics[width=\textwidth]{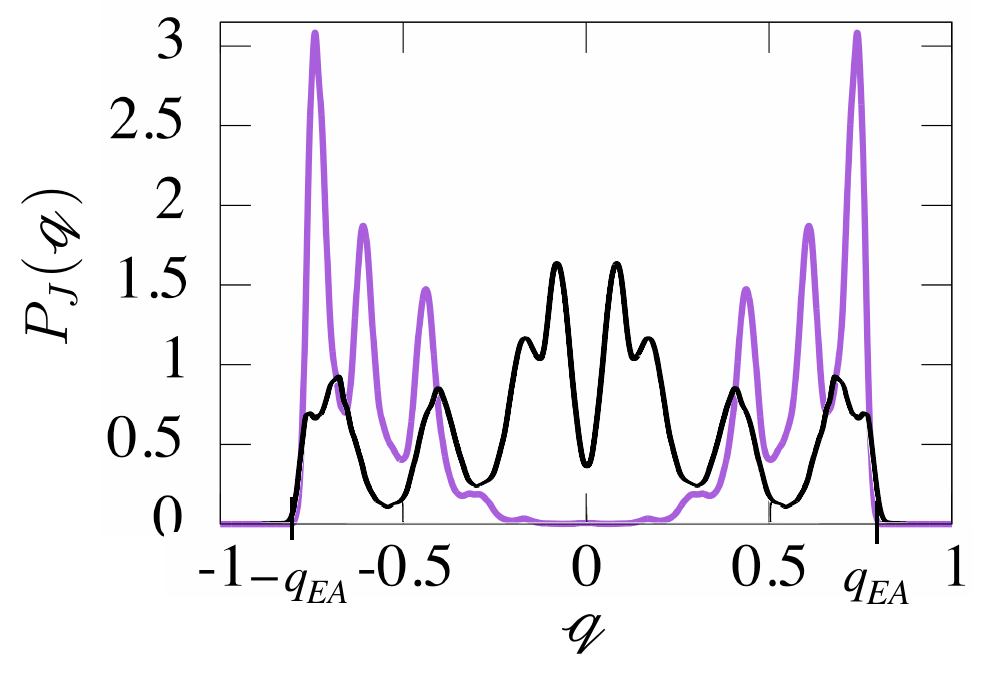}
         %\caption{}
         \label{fig:pj_variation}
     \end{subfigure}
         % \begin{subfigure}[b]{0.45\textwidth}
         %\centering
         %\includegraphics[width=\textwidth]{figures/dr_e_snapshot_distribution_N1024_l0004_wc0006267_n51_equil_50_X_qw1_gamma0000003_gammap0000003_eminscratch_1nice_.png}
         %\caption{ }
         %\label{fig:pRq}
     %\end{subfigure}
     \caption{%Comparison of the order parameter distribution of a spin glass and the polarization distribution in a polarization glass. \textbf{Left:} 
     Example of two rough probability distributions $P_J(\mathscr{q})$ of the overlap $\mathscr{q}$  for an Edwards-Anderson model of a spin glass without external magnetic field (aggregated from Ref. \citenum{parisi_nobel_2023} based on data from Refs. \citenum{alvarez_banos_nature_2010,alvarez_banos_static_2010}). The different realizations of $J$ show large deviations even in the thermodynamic limit. %A smooth curve $P(q)$ is only achieved after averaging over many realizations of $J$ (see e.g. Fig. \ref{fig:aging}). 
     %\textbf{Right}: Likewise, two (discretely) distinguishable polarization distribution are arising by individually converging the cH equations for two different realizations of parameter sets $(R,q_\beta=0)$ (Shin-Metiu setup with $N=1024$ molecules). 
     }
     \label{fig:pdistr}
\end{figure}

\begin{figure}
         \centering
         \includegraphics[width=\textwidth]{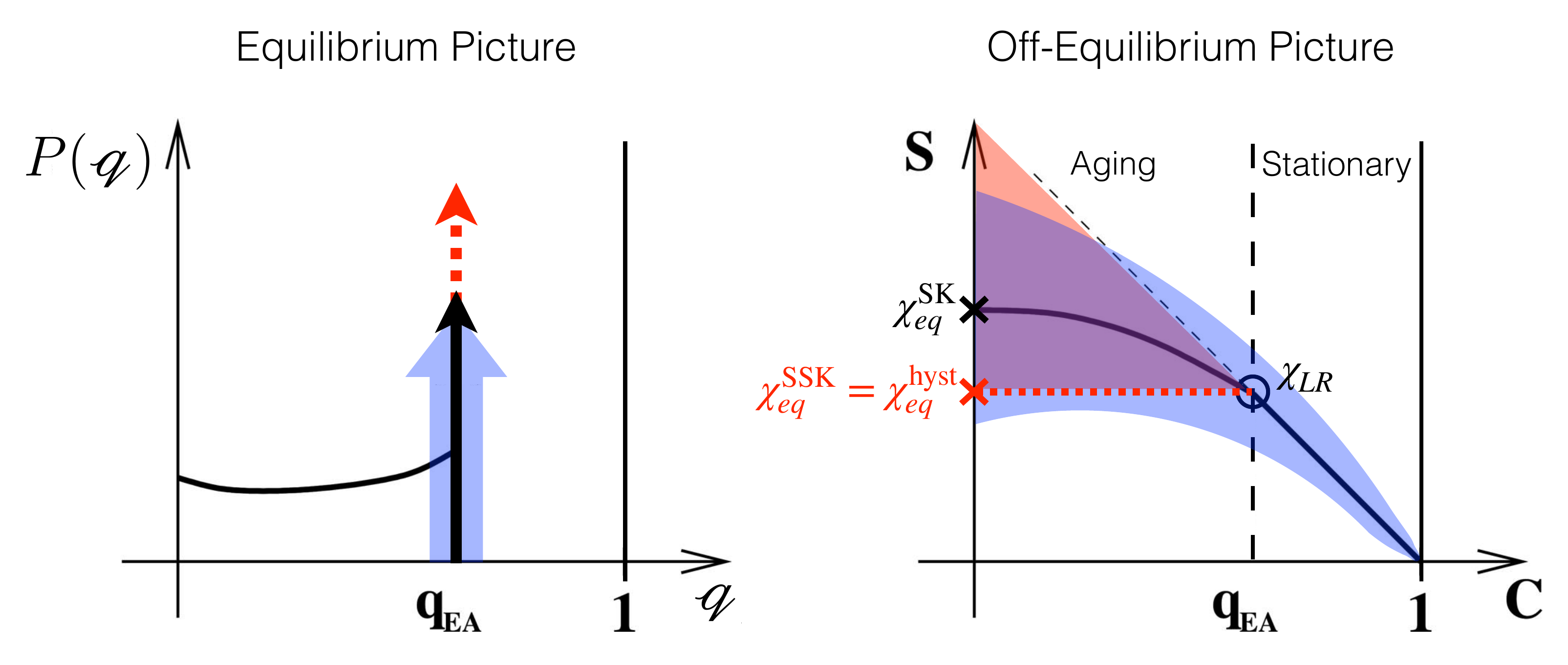}
         \caption{Connecting equilibrium and off-equilibrium picture for the SK (black) and SSK (red) spin glass assuming identical $q_{\rm EA}$. \textbf{Left}:  Equilibrium order parameter distribution  of a SK model (black) adapted from Ref. \citenum{parisi_generalized_1999} in comparison with deviations of $q_{EA}(t)$ in a cavity.
         \textbf{Right}: In spin glasses two different (weak) off-equilibrium regimes (stationary and aging) can be distinguished for a small external field perturbation $h^\prime$,  applied after a waiting time $t_w$. The stationary regime is governed by a linear fluctuation dissipation  relation, which terminates at $C=q_{EA}$ given by the linear response susceptibility $S(q_{EA})=T \chi_{LR}$.  In contrast, the aging regime is governed by the modified fluctuation dissipation relations in Eq. \eqref{eq:dSdC}. It is bounded  by the full thermal equilibrium susceptibility  $S(0)=T\chi_{eq}$ at $C=0$. Notice that $\chi_{eq}\neq  \chi_{LR}$  indicates replica symmetry breaking, whereas $\chi_{eq}=  \chi_{LR}$ usually indicates normal hysteresis effects (red) that do not require the complex theory of spin glasses to  describe its equilibrium properties. At first sight, the SSK model appears identical, i.e., trivial, in that picture. Nevertheless, its dynamics remains highly non-trivial, since it depends asymptotically on the waiting time according to Eq. \eqref{eq:ssk_asym}. The light-blue region indicates that cavity-mediated correlations will deviate from an ideal SSK model, since the underlying probability distribution obviously depends on the chemical system and cannot be considered normally distributed. Corresponding investigation will require expensive numerical simulations.}
         \label{fig:aging}
     \end{figure}

In order to better understand physical consequences of the SK model defined in Eq.~\eqref{eq:sk_model}, we have a closer look at the average free energy $F$, which can be defined as follows in the thermodynamic limit $N\rightarrow\infty$,~\cite{parisi_nobel_2023}  
\begin{eqnarray}
    F(T)&=& - \lim_{N\rightarrow\infty}\frac{\overline{k_B T\log(Z_J(T,N))}}{N},\\
    Z_J(T,N)&=& 2^{-N}\sum_{\{\boldsymbol{\sigma}\}}e^{-\frac{H_J(\boldsymbol{\sigma})}{k_B T}}.
\end{eqnarray}
The overline indicates the averaging of the partition function $Z_J$ with respect to randomly drawn $J_{ij}$ realizations. To simplify the $J$-averaging, the so-called replica trick was proposed, where instead of one system with $N$-spins, an extended system consisting of $n$-times the same system made of $N$-spins is considered. It simplifies taking the logarithm as follows~\cite{parisi_nobel_2023}
\begin{eqnarray}
    F_n(T)&=& - \lim_{N\rightarrow\infty}\frac{\overline{k_B T (Z_J(T,N))^n}}{nN},\\
    F(T)&=&\lim_{n\rightarrow 0}F_n(T).
\end{eqnarray}
The replica trick works if $F_n$ is analytic in $n$ and has no singularities.~\cite{parisi_nobel_2023} In particular, one would expect replica symmetry to hold as a natural assumption, since it implies the re-shuffling of the $n$ identical replicas will not change the result. From this symmetry aspect, one can deduce that the free energy depends on a single order parameter $q$, which then can be minimized to determine the alleged solution of the SK model.~\cite{sherrington_solvable_1975, parisi_nobel_2023} However, it turns out that $F_n$ is indeed not analytic for $n<n_c<1$ in the SK model. This indicates that the replica symmetry is spontaneously broken and thus things become much more complex.
After a long endeavour the exact solution of the replica Ansatz was discovered by Parisi yielding the following free energy~\cite{parisi_infinite_1979, parisi_sequence_1980}
\begin{eqnarray}
    F&=& \max_{\mathscr{q}(\mathscr{x})} F[\mathscr{q}(\mathscr{x})],
\end{eqnarray}
where the corresponding partial differential equations are given explicitly in footnote~\footnote{
\begin{eqnarray}
    F[\mathscr{q}(\mathscr{x})]&=& -\frac{1}{4 k_B T}\bigg[1+\int _0^1 dx \mathscr{q}^2(\mathscr{x})-2 \mathscr{q}(1)\bigg]-k_B T f(0,0)\nonumber\\
    \frac{\partial f(x,h)}{\partial \mathscr{x}}&=& -\frac{1}{2}\bigg[\frac{\partial^2 f}{\partial h^2}+ x \bigg(\frac{\partial f}{\partial h}\bigg)^2\bigg],\nonumber
\end{eqnarray}
with $f(1,h)=\ln(2 \cosh(h/k_B T))$. }. 
Considerably later it was also proven that the exact solution of the replica ansatz  indeed determines the exact free energy of the original problem.~\cite{guerra_broken_2003,talagrand_parisi_2006} For the subsequent discussions this solution will be of minor relevance. However, the emergence of an order function $\mathscr{q}(\mathscr{x})$, instead of just an order parameter $\mathscr{q}$, is essential for the mechanistic understanding of (many) spin glasses.
In more detail, explicit expressions were found that relate the order parameter function $\mathscr{q}(\mathscr{x})$ to the probability density 
\begin{eqnarray}
     P_J(\mathscr{q})=\sum_{\alpha \gamma} w_\alpha w_\gamma
\delta(\mathcal{Q}_{\alpha\gamma}-\mathscr{q}), \label{eq:pjq}
\end{eqnarray} 
of finding two states with overlap
 \begin{eqnarray}
     \mathcal{Q}_{\alpha\gamma}=\frac{\sum_i m(i)_\alpha m(i)_\gamma}{N}
 \end{eqnarray} 
 in a given sample $J$. The statistical weights of solution $\alpha$ are indicated by $w_\alpha$.~\cite{parisi_spin_2006}  Notice the connection of $\mathcal{Q}_{\alpha\gamma}$ to the Edwards-Anderson order parameter or self-overlap $\mathcal{Q}_{\alpha\alpha}=q_{EA}$. 
An astonishing feature of spin glasses in general is that  $P_J(\mathscr{q})$ shows a dramatic dependency on the specific choice of $J$ even in the thermodynamic limit (see Fig.~\ref{fig:pdistr} for the Edwards-Anderson model without external magnetization fields). 
A smooth curve is only achieved by averaging over all possible realizations $J$ (see, e.g., Fig.~\ref{fig:aging}) yielding the equilibrium overlap
\begin{eqnarray}
    P(\mathscr{q})=\overline{P_J(\mathscr{q})}.
\end{eqnarray}
Eventually, the functional dependency of $\mathscr{q}(\mathscr{x})$ can be made explicit by inverting  the probability density
\begin{eqnarray}
    \mathscr{x}(\mathscr{q})=\int _0 ^\mathscr{q} d\mathscr{q}^\prime P(\mathscr{q}^\prime).
\end{eqnarray}
Notice, a signature of replica symmetry breaking~\cite{parisi_nobel_2023} is the deviation of $P(\mathscr{q})$ from \underline{two} delta functions at $\mathscr{q}=\pm q_{EA}$ as illustrated in Fig.~\ref{fig:aging}.

\subsubsection{Equilibrium Susceptibilities}
From an experimental point of view, a characterization of a spin glass is usually done by varying the temperature and applying external magnetization-field  perturbations $h^\prime$ as
\begin{eqnarray}
    H^\prime(\boldsymbol{\sigma}) =  h^\prime\sum_i \sigma_i.
\end{eqnarray} 
In our polaritonc picture, this is equivalent to changing the (dressed electronic) temperature or applying a small external electric field perturbation.% $E_{\rm ext}$, i.e., the Hamiltonian is applicable \revMR{(I do not understand)} and is perturbed by,where $E_{\rm ext}$ determines $h^\prime=$ \revDS{define explicitly? will contain direction of molecules}.

\begin{figure}
\begin{subfigure}[b]{\textwidth}
 \centering
         \includegraphics[width=0.6\textwidth]{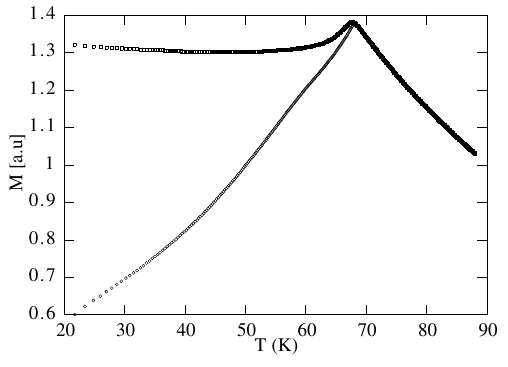}
        \end{subfigure}
        %\begin{subfigure}[b]{\textwidth}
        % \centering
        % \includegraphics[width=0.6\textwidth]{susceptibility_theory.pdf}
         %\caption{}
        % \end{subfigure}
         \caption{Experimental susceptibilities with respect to temperature $T$. The linear response susceptibility ($\chi_{LR}$, lower curve) can be measured experimentally by applying an external field perturbation after the cooling of the material (Cu(Mn13.5\%)) below the critical spin glass temperature. In contrast, the equilibrium susceptibility ($\chi_{eq}$, upper curve) can be approximated experimentally by applying the magnetic field perturbation before the cooling. \cite{djurberg_magnetic_1999,parisi_spin_2006} Above the critical temperature the material enters the paramagnetic stable phase, with only one susceptibility, i.e., the linear-response accesses equilibrium properties.}
         \label{fig:suscept}
\end{figure}

The hallmark of replica symmetry breaking in spin glass theory can be attributed to the emergence of two different static equilibrium susceptibilities,~\cite{parisi_spin_2006} which are observed in experiments~\cite{djurberg_magnetic_1999} and in the SK model.~\cite{mezard_spin_1987} The two extreme cases describe the response of the system subject to a small external field perturbation. In the so-called zero-field cooled case, the system remains inside a given state while changing the magnetization (electric, for the polaritonic setup) field, with a corresponding linear response susceptibility $\chi_{LR}$. In contrast, the (true) thermodynamic equilibrium susceptibility $\chi_{eq}$ describes the situation, where the spin glass is allowed to relax to the thermodynamically most favored state in presence of a weak external field perturbation. 
\begin{eqnarray}
    \chi_{LR} = \frac{1-q_{EA}}{k_B T}\label{eq:chilr}\\
    \chi_{eq} = \frac{\int dx (1-\mathscr{q}(\mathscr{x}))}{k_B T}
\end{eqnarray}
Experimentally, the static linear-response susceptibility $\chi_{LR}$ can be measured by looking at the response to a small external field perturbation $h^\prime$, after cooling the system to the desired low temperature. In contrast, the equilibrium susceptibility $\chi_{eq}$ is approximated by the field cooled susceptibility, which is measured by applying the small field perturbation already while cooling the system below the spin glass transition temperature. In this case, the system has time to explore and select the most appropriate state while cooling in presence of the external perturbation. 
Experimental results of the two different spin glass susceptibilities below the critical temperature are  illustrated for Cu(Mn13.5\%) in Fig.~\ref{fig:suscept} %and compared with theoretical predictions
.~\cite{djurberg_magnetic_1999,parisi_spin_2006}
In spin glass theory, e.g., the SK model, there is a clear distinction between replica symmetry breaking and hysteresis effects:~\cite{parisi_spin_2006} Hysteresis is commonly attributed to defects that are localized in space and induce a finite free-energy barrier and thus finite lifetime of meta-stable states. Thus, in hysteresis, the two susceptibilities coincide after waiting sufficiently long time. In contrast, the non-local barriers in spin glasses imply the re-arrangements in arbitrary large regions of the system, which can even diverge in the thermodynamic limit. Therefore, the different susceptibilities will not agree, provided that the externally applied field perturbation is sufficiently small. 

A clear distinction between hysteresis and a spin glass  is not so trivial in a polaritonic setup, as we will see in Sec.~\ref{sec:ssk}, and thus requires more experimental and theoretical work.~\cite{schwartz_erc_2024} Moreover, as we can deduce from the previous discussions, the quasi-static spin glass picture of the SK model is incomplete for our polaritonic setup, since the dressed  electronic-structure of an ensemble of molecules under VSC will be (periodically) driven by the dynamics of the nuclear and displacement field coordinates. Therefore, it is likely that time-dependent external field perturbations are required to probe the coexistence of different cavity-induced linear (or higher-order) susceptibilities. In particular, we already see that Eq.~\eqref{eq:chilr} should explicitly depend on time, since $J_{ij}$ and thus the Edwards-Anderson order parameter  depend on $(R(t),q_\beta(t))$ and thus $q_{EA}(t)$. If this also implies the coexistence of different (possibly dynamic) susceptibilities under VSC remains a non-trivial open question. A further hint at interesting effects is a highly degenerate ground state, which makes response calculations intricate. However, investigating dynamic susceptibilities could provide a promising route to verify and characterize the proposed polarization-glass phase in a cavity.

\subsubsection{Off-equilibrium spin glass: Breakdown of the fluctuation-dissipation theorem and aging effects\label{sec:offequil}}

Even more interesting than the equilibrium properties of spin glasses are their associated off-equilibrium  phenomena, which are observable for different materials. 
%The response of a spin glass to an externally applied field perturbation, which deviates from its initially perturbed state (off-equilibrium), can be characterized in terms of the overlap $q$  with  respect to the initial state, i.e,\cite{parisi_nobel_2023}
%\begin{eqnarray}
%  \chi(q)=\frac{1}{k_B T}  \int_q^1 dq^\prime x(q^\prime).
%\end{eqnarray}
%The two extreme cases then correspond to the previously discussed equilibrium picture, i.e., $\chi_{LR}=\chi(q_{EA})$ and $\chi_{eq}=\chi(0)$. 
In order to measure the generalized susceptibility, it is common practice to rely on the fluctuation-dissipation theorem, which relates the response of the system to an external perturbation (weak off-equilibrium) to its equilibrium properties (fluctuations/correlations). The fluctuations of glassy systems can be characterized by the time-correlations of the magnetizations (or delocalized/intermolecular excitations and their induced polarizations, respectively, for the polaritonic setup), 
\begin{eqnarray}
    C(t,t_w)= %\langle m(t_w)m(t_w+t)\rangle\label{eq:autocorr}=
    \frac{1}{N}\sum_i^N \langle \sigma_i(t_w) \sigma_i(t_w+t)\rangle 
    %\Rightarrow\frac{1}{N}\sum_i^N \frac{\langle \Delta \mu_i(t_w) \Delta \mu_i(t_w+t)\rangle}{\langle \Delta \mu_i(t_w) \Delta \mu_i(t_w)\rangle}.
    \label{eq:corr}
\end{eqnarray}
%Comparing the magnetization time-correlations  of a (quasi-static) spin glass with the electronic correlations of a polarization glass reveals similarities (see Fig.~\ref{fig:correlations}) but again highlights some fundamental physical differences.
In a spin glass, the magnetization correlations decay monotonically, but extremely slowly even on a logarithmic time-scale. In addition, the random process of $\sigma_i(t)$ cannot be considered a wide-sense stationary stochastic process, which means the correlation does not only depend on the time-difference $t$, but also on the waiting-time $t_w$ elapsed since entering the spin glass phase. In other words, time-reversal symmetry is explicitly broken in a spin glass.  The explicit waiting-time dependency will give rise to specific \textbf{aging effects} that are discussed in in the following: %In contrast, the polarization glass correlations reveal different oscillatory regimes, introduced by the time-dependent $(R(t),q_\beta(t))$-parameters. MD simulations with $N=36$ Shin-Metiu molecules reveal the following correlation features within an adiabatic regime:

\begin{figure}
     \begin{subfigure}[b]{\textwidth}
         \centering
         \includegraphics[width=\textwidth]{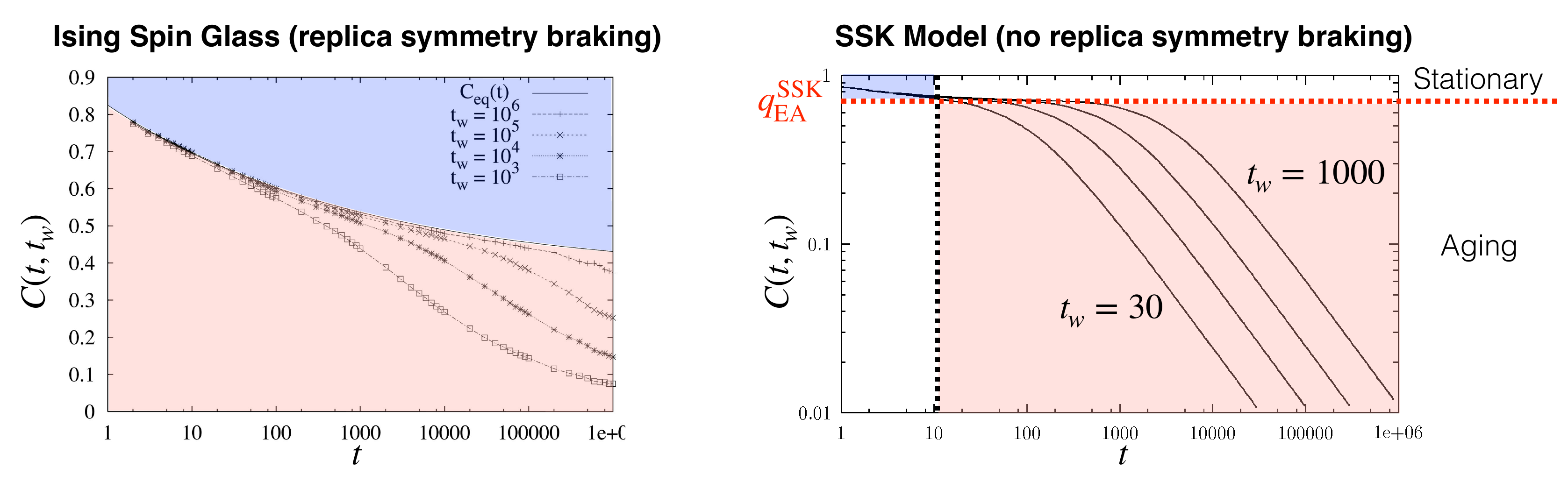}
        %\caption{}
         \label{fig:corr_sg}
     \end{subfigure}
         % \begin{subfigure}[b]{\textwidth}
         %\centering
         %\includegraphics[width=0.6\textwidth]{dr_e_autocorrelation_comparison_N36.png}
         %\caption{}
         %\label{fig:corr_cH}
     %\end{subfigure}
     \caption{%Comparison of time-correaltions in a spin and polarization glass: \textbf{Top:} 
     Comparison of aging behavior for two spin glass models with (left) and without (right) replica symmetry breaking. Stationary correlation regime is marked in light blue and the aging regime in light red. Left: Monotonically decaying time correlation functions $C(t,t_w)$ of an Ising spin glass (taken from Ref.~\citenum{parisi_spin_2006} based on data from Ref. \citenum{marinari_violation_1998}). The dependency on the waiting time $t_w$ indicates the aging behavior, i.e., the deviation from $C_{\rm eq}(t)$. Right: Time-correlation functions of the Spherical Sherrington Kirkpatrick (SSK) model. Notice the plateau region of the autocorrelation function at $C=q_{\rm EA}$, which depends asymptotically on the waiting time, thus the correlations may not necessarily decay to zero. Data taken from Ref. \citenum{cugliandolo_full_1995}.%\textbf{Bottom:} Oscillating polarization correlations of the Shin-Metiu setup under VSC with $N=36$ molecules. When calculating $C(t)$ it was implicitly assumed that $\Delta \mu_i(t)$ is a wide-sense stationary stochastic process within the adiabatic cH picture. The correlations oscillate fast (light-blue), on the time-scale of the cavity (vibrations), where the bold lines visualize the enveloping function. The blue line corresponds to the strong coupling to one lossy cavity mode ($q_\beta$ weakly coupled to a Langevin bath). In contrast,  the coupling to a  Brownian mode is shown in red, which can be interpreted as the weak coupling to many cavity modes (thermal bath) instead.   }
     }
     \label{fig:correlations}
\end{figure}

For simplicity, let us make the following assumptions: At time $t_0=0$ the system is suddenly cooled below the critical glass temperature $T<T_g$, which triggers the phase transition into a spin glass. After a waiting time $t_w$, a constant external field perturbation $h^\prime$ is applied. Eventually, the response of the system $S(t_w,t)$ is evaluated at time $t+t_w\geq t_w$.      
Under these conditions, the fluctuation-dissipation relations connect~\cite{cugliandolo_out--equilibrium_1994,franz_quasi-equilibrium_2000,parisi_spin_2006}  the time correlation function of the total magnetization $C(t,t_w)$, defined in Eq.~\eqref{eq:corr},
to the average relaxation function per spin,  which is defined by~\cite{parisi_spin_2006} 
\begin{eqnarray}
    S(t,t_w)=k_B T \lim_{\delta h^\prime\rightarrow 0}\frac{\delta \langle m(t_w+t) \rangle}{\delta h^\prime}.%=k_B T\chi(t_w,t).
    \label{eq:sdef}
\end{eqnarray}
Eq.~\eqref{eq:sdef} describes the response of the magnetization at time $t_w+t$ if the external field was added at time $t_w$. 
However, the standard fluctuation-dissipation theorem is only applicable if the system obeys the detailed balance condition,~\cite{kubo_statistical_1991} which can be violated in a glassy system on longer time-scales (e.g., by spontaneous replica symmetry breaking).
To account for this aspect, modified fluctuation dissipation relations have been proposed that hold for weak off-equilibrium .~\cite{cugliandolo_analytical_1993,franz_measuring_1998,parisi_nobel_2023} 
In more detail, two different off-equilibrium regimes are distinguished for (spin) glasses, for which different fluctuation-dissipation relations hold:~\cite{franz_quasi-equilibrium_2000} %the \textit{stationary correlation regime} and the \textit{aging regime}. 

\begin{figure}
         \centering
         \includegraphics[width=0.5\textwidth]{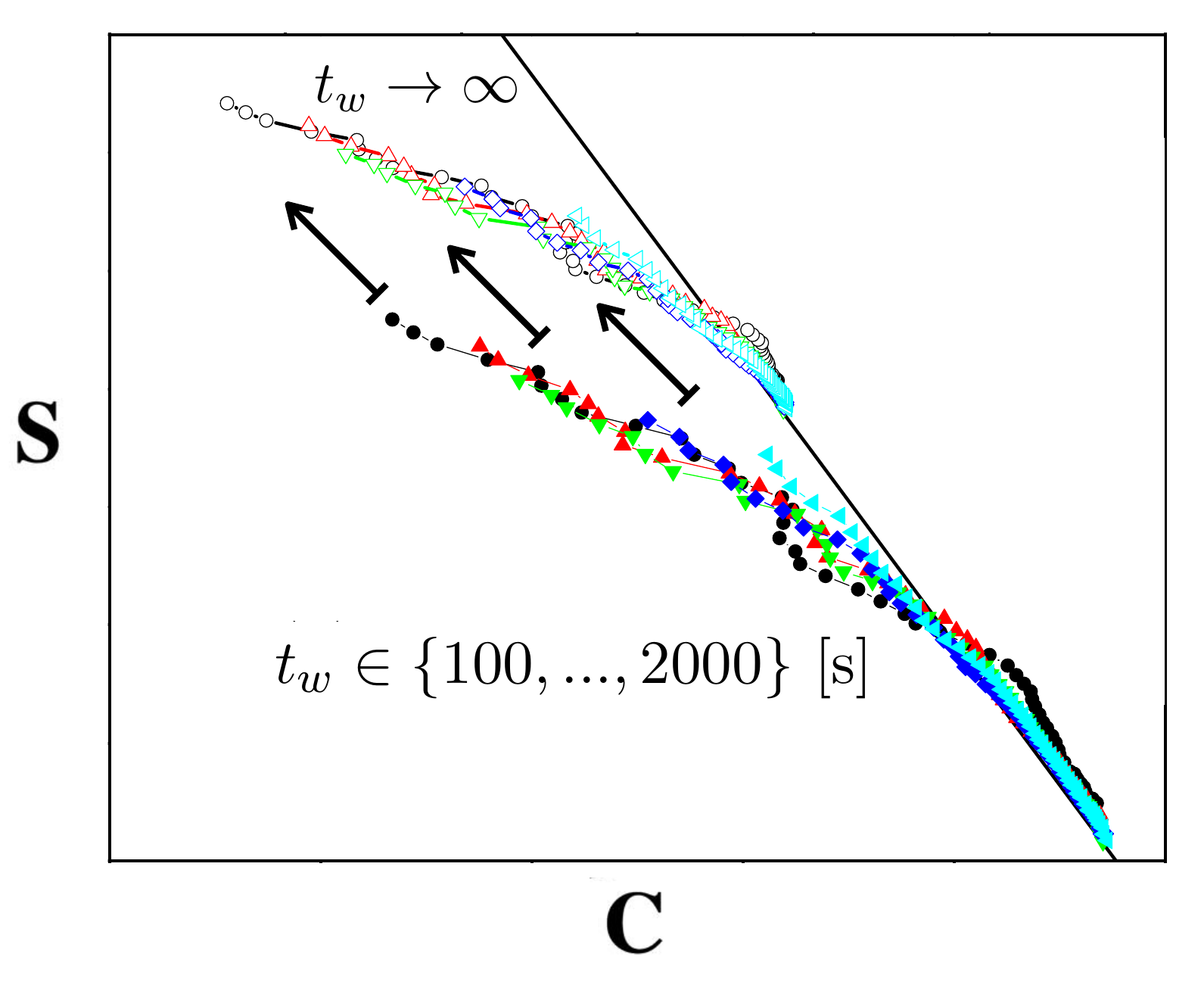}
         \caption{ Experimental off-equilibrium measurement: The breakdown of fluctuation-dissipation relations (aging) measured the after cooling CdCr$_{1.7}$In$_{0.3}$S$_4$ below the critical spin glass temperature of $T_g=16.2$ K. Bold symbols indicate the measured relaxation-correlation curve for different (finite) waiting times $t_w \in \{100, 200, 500, 1000, 2000\}$ [s]. Open symbols show the extrapolation to infinite waiting times $t_w\rightarrow\infty$ .%for which the modified fluctuation dissipation relations in Eq. \eqref{eq:dSdC} are directly related to equilibrium properties of the spin glass (see e.g. Eq. \eqref{eq:agingvseq}).  
         The illustration was modified based on Fig. 7 Ref. \citenum{parisi_nobel_2023}, which contains experimental data from Ref. \citenum{herisson_fluctuation-dissipation_2002}.}
         \label{fig:aging_expt}
     \end{figure}

In the \textit{stationary correlation regime} of a spin glass, the correlations are assumed to solely depend on $t$, but not on the waiting time $t_w$, i.e, $C(t,t_w)=C_s(t)$. Typically this approximation is reasonable only for relatively small $t\approx 0$  in spin glasses, which implies high correlations $C_s(t)\approx 1$. Having correlations close to unity, the standard (thermal equilibrium) fluctuation-dissipation relation are applicable, which yields~\cite{marinari_violation_1998,baity-jesi_statics-dynamics_2017}
\begin{eqnarray}
 S(t)=1-C_s(t).   \label{eq:eq_flucrel}
\end{eqnarray} 
In the \textit{aging regime}, where the correlations are no longer stationary, modified fluctuation-dissipation relations of the following form were suggested~\cite{parisi_nobel_2023}
\begin{eqnarray}
    \frac{d S(t_w,t) }{dt}=%\frac{1}{k_B T} 
    X(C(t,t_w))\frac{d C(t,t_w)}{dt}.
    \label{eq:dSdC}
\end{eqnarray}
Again the SK model provides an ideal starting point to interpret Eq.~\eqref{eq:dSdC} analytically, since one can show that $ C(t,t_w)=\frac{1}{N}\sum_i^N \langle \sigma_i(t_w) \rangle\langle\sigma_i(t_w+t)\rangle=q(t_w,t_w+t)$ in absence of an external magnetization field, i.e., for $h_m=J_0=0$.~\cite{parisi_spin_2006} 
This allows to discuss off-equilibrium effects in spin glasses analytically, in the absence of external magnetization fields. In more detail, by eliminating the time parametrically, one can re-express $S(t_w,t)\mapsto S(t_w,C)$. Afterwards, identifying $X(C)= d S(t_w,C)/d C$ in the large waiting limit $t_w\rightarrow\infty$, one can relate the dynamic quantity $X(C)$ to the equilibrium $\mathscr{x}(\mathscr{q})$, i.e.,  $X(C)=\mathscr{s}(\mathscr{q})|_{\mathscr{q}=C}$.~\cite{parisi_nobel_2023}
This leads to a simple physical picture in the aging regime in terms of the slope of the response with respect to the correlations, i.e.,\cite{franz_off-equilibrium_1994, parisi_spin_2006, cugliandolo_analytical_1993, cugliandolo_energy_1997, cugliandolo_out--equilibrium_1994, franz_response_1999}
\begin{eqnarray}
    \frac{dS}{dC}=X(C)=\int_0 ^C d\mathscr{q} P(\mathscr{q}).\label{eq:agingvseq}
\end{eqnarray}
In other words, the deviations  of the fluctuation-dissipation theorem that are caused by aging effects can be related to equilibrium properties given by $P(\mathscr{q})$. An illustration of the two different off-equilibrium regimes with their relation to equilibrium properties is given in Fig.~\ref{fig:aging} for the SK model in comparison with standard hysteresis effects, i.e., visualizing the difference between replica symmetry breaking and hysteresis. Note that the modifications of the fluctuation-dissipation relations can be re-interpreted in terms of an effective temperature~\cite{cugliandolo_energy_1997, franz_measuring_1998}
\begin{eqnarray}
\tau=-T   \bigg(\frac{dS}{dC}  \bigg)^{-1}\geq T\label{eq:effT}
\end{eqnarray}
which indicates a heating or excess of thermal fluctuations since $0\leq dS/dC \leq 1$ according to the probability interpretation of Eq.~\eqref{eq:agingvseq}.  The two different off-equilibrium regimes, i.e., the emergence of aging effects, seem to be a generic feature of glassy systems.~\cite{parisi_nobel_2023} In Fig.~\ref{fig:aging_expt} experimentally recorded fluctuation-dissipation relations are shown for CdCr$_{1.7}$In$_{0.3}$S$_4$ with respect to different finite waiting times.~\cite{herisson_fluctuation-dissipation_2002} %The extrapolation to infinite waiting times allows the connection to the theoretical spin glass models (i.e., the connecting to equilibrium properties, except for the ). 

%Including the parametric feedback of the nuclei on the dressed  electronic-structure modifies the fluctuation-dissipation picture of a spin glass significantly. 
%In particular, the emergence of periodic oscillations implies that the thermal equilibrium fluctuation-dissipation relation, given in Eq.~\eqref{eq:eq_flucrel}, will no longer be applicable in the stationary correlation picture. This is not so surprising since the electronic subsystem is strongly correlated with the other subsystems (not decoupled, i.e. strongly hybridised). Therefore, Eq.~\eqref{eq:eq_flucrel} will remain valid solely for very short time-scales ($C_s(t)\approx 1$), or if the cavity is too lossy/too weakly coupled to the system to induce any chemical changes on the molecular ensemble (see suppressed correlations visualized by the red line in Fig.~\ref{fig:correlations}). Consequently, different fluctuation-dissipation relations are expected even in the stationary correlation regime under VSC. Similarly, by increasing the collective coupling strength further, the breakdown of the adiabatic approximation suggest the emergence of a dynamic aging regime under VSC. In analogy to the stationary case, we expect that the dynamic feedback effects require modifications of the fluctuation-dissipation relation in Eq.~\eqref{eq:dSdC} to be captured. \revDS{update and refer to discussion of spherical spin glass... possibly move parts of interpretation there.}

\subsection{Spherical Sherrington-Kirkpatrick (SSK) model\label{sec:ssk}}

The spherical Sherrington-Kirkpatrick (SSK) model (or spherical 2-spin glass) closely resembles the SK model introduced in Eq. \eqref{eq:sk_model}. However, the SSK model possesses continuous spin variables $s_i$, which additionally obey the following ``spherical'' constraint $\sum_{i }s_i^2=N_J$ that ensures a finite ground state energy.  The SSK model was introduced by Kosterlitz, Thouless, and Jones in Ref.~\citenum{kosterlitz_spherical_1976} as
\begin{eqnarray}
    H_{SSK}(\boldsymbol{s}) =-\sum_{i<j}^{N_J}s_i s_j J_{ij},\ J\sim\mathcal{N}(J_0/N_J,\tilde{J}^2/N_J).\label{eq:ssk}
\end{eqnarray}
For simplicity, we assume a symmetric random matrix, i.e, $J_{ij}=J_{ji}$ in the following, as it is the case for the DSE correlation energy. 
In contrast to the SK model, the SSK model is easier to analyze\cite{baik_fluctuations_2016}. In particular, a simple analytic solution for the free-energy was found using random-matrix theory.\cite{kosterlitz_spherical_1976} 
In the absence of an external magnetic field, the expected free energy per site (averaged over different realizations of $J_{ij}$) is given  as\cite{kosterlitz_spherical_1976} 
%\revDS{Ruggi: I think one needs to replace $T\mapsto k_B T$. do you agree?} \revMR{(In other papers, e.g., Talagrand https://www.jstor.org/stable/25450054, the F is defined with a minus sign yet the signs in the formula below are not just swapped. Also in https://link.aps.org/accepted/10.1103/PhysRevE.85.041127, that seems to have the same sign convention, we get something slightly different. I do not fully understand. But at least in Talagrand they us $\beta$ and if we use $\beta =1/(k_bT)$ we get something similar)}
\begin{eqnarray}
     F_{T,\tilde{J}}= \lim_{N_J\rightarrow\infty}F_{N_J,T,\tilde{J}}=\begin{cases}
-\frac{\tilde{J}^2}{4k_{B}T}-\frac{1}{2}k_{B}T(1+\ln(2)) & \text{if } T \geq T_c,\ J_0=0 \\
\frac{1}{2}k_{B}T\ln\big(k_{B}T/(2\tilde{J})\big)-\tilde{J}+\frac{1}{4}k_{B}T & \text{if } T <T_c,\ J_0=0,
\end{cases}
\end{eqnarray}
in the thermodynamic limit. The Edwards-Anderson order parameter is given by $q_{EA}^{\rm SSK}=1-T/T_c$.
It reveals a spin glass phase for temperatures below the critical temperature $T_c=\tilde{J}/k_{B}$. From a static point of view the model is relatively simple, since  the SSK model does not break replica symmetry.\cite{cugliandolo_full_1995} In particular, its energy has only two minima,\cite{cugliandolo_full_1995} which suggests the absence of frustration and thus trivial dynamics. However, it turns out that the energy-fluctuations~\cite{baik_fluctuations_2016} and the dynamics~\cite{cugliandolo_full_1995} of the SSK model are modified non-trivially. Indeed, significant aging effects appear and for almost any initial condition, the glassy system is out of equilibrium, and the evolution does not (!) lead to equilibrium.~\cite{cugliandolo_full_1995}.

In more detail, the order (decaying with respect to the number of spins) and the distribution of the free-energy fluctuations per spin among the ensemble changes as follows, 
\begin{eqnarray}
     F_{T,1}-F_{N_J,T,1}\propto\begin{cases}
\frac{1}{N_J} \mathcal{N} & \text{if } T \geq T_c,\ J_0=0 \\
\frac{1}{N_J^{2/3}}TW_1  & \text{if } T <T_c,\ J_0=0,
\end{cases}\label{eq:freefluct}
\end{eqnarray}
i.e., locally, when assuming $\tilde{J}=1$.~\cite{baik_fluctuations_2016}
This means, within the spin glass phase ($T<T_c$) the (local) free-energy fluctuations are increased (i.e., decay as $N_J^{-2/3}$ instead of $N_J^{-1}$). Moreover, their probability distribution changes from a normal distribution $\mathcal{N}$ to a Gaussian orthogonal ensemble Tracy-Widom distribution $TW_1$.~\cite{baik_fluctuations_2016} The Tracy-Widom probability distribution is skewed to the right and decays slower than a gaussian for positive values, i.e., possesses a heavier right tail.\cite{majumdar_top_2014} Consequently,  the mean of the fluctuations is only moderately affected (bulk dispersion forces) by the skewness, whereas rare events (distribution tails) are expected to  change substantially below the critical temperature.   

For the SSK model, analytic solutions can be found for the two-time spin correlation function defined in Eq.~\eqref{eq:corr} in the spin glass phase.~\cite{cugliandolo_full_1995} For this purpose, Cugliandolo and Dean propagated the spin dynamics (defined via Langevin equations) for different initial conditions. Surprisingly, for almost any initial condition non-equilibrium dynamics occurs! Only for very specific initial conditions, with ``a macroscopic condensation" determined by the eigenvalue of the random matrix $J_{ij}$, equilibrium dynamics is achieved.~\cite{cugliandolo_full_1995} In the following, we focus on the spin correlations of the generic non-equilibrium setup, for which we recover the stationary correlation regime for relatively small $t$ and the aging regime, where the correlations explicitly depend on the waiting time $t_w$ (see Fig. \ref{fig:correlations}). The \textit{stationary correlation regime} 
\begin{eqnarray}
    C(t,t_w)\sim
 C_s(t), \  t\ll t_w.
\end{eqnarray}
is characterized by a fast initial decay from $C=1$ to $C=q_{EA}$ for $t\ll t_w$, where the usual fluctuation dissipation relations hold according to Eq. \eqref{eq:eq_flucrel}. 
The \textit{aging regime} can be subdivided into two further sub-regimes as follows,
\begin{eqnarray}
     C(t,t_w)\sim\begin{cases}
q_{EA}^{\rm SSK}=1-T/T_c &  \text{if } t \approx t_w, \\
\rm{slowly\ decaying \ to\ 0} &\text{if } t \gg t_w .
\end{cases}\label{eq:agingsub}
\end{eqnarray}
One of the striking features of the SSK model is that the plateau region of the autocorrelation function in Eq. \eqref{eq:agingsub}, depends asymptotically on the waiting time, i.e., \cite{cugliandolo_full_1995}
\begin{eqnarray}
    \lim_{t_w\rightarrow\infty}C(t,t_w)=q_{EA},\ \forall t \ \rm{finite}.\label{eq:ssk_asym}
\end{eqnarray} 
Furthermore, the decay of the correlation for fixed $t\gg t_w$ is extremely slow (notice the logarithmic time-scale in Fig. \ref{fig:correlations}). 

However, despite above long-lived correlations, aging effects in the magnetization are weak, i.e., the relaxation of an externally induced magnetization $S$ defined in Eq. \eqref{eq:sdef} decays fast. As a consequence, one can show that the modified fluctuation-dissipation relation are sort of trivial and one finds \cite{cugliandolo_full_1995}
\begin{eqnarray}
    \frac{dS}{dC}=X(C)\sim\begin{cases}
-1 &  \text{if}\  C(t,t_w)\sim
 C_s(t),  \\
0 &\text{otherwise, }\label{eq:modflucdis}
\end{cases}
\end{eqnarray}
as visualized in Fig. \ref{fig:aging}. The condition $\frac{dS}{dC}=0$ suggest a divergent effective temperature $\tau\rightarrow\infty$, i.e., a strong excess of thermal fluctuations in the aging regime, which we interpret as a sign of the persistent non-equilibrium dynamics of the SSK model. At the same time, the linear-response susceptibility matches the equilibrium susceptibility $\chi_{LR}^{\rm SSK}=\chi_{eq}^{\rm SSK}$, similar to standard hysteresis effects. Thus applying a small static external field perturbation does not probe the complex correlations of the SSK model. As previously mentioned, applying time-dependent external fields may be an interesting option for VSC. However, at the moment it remains unclear if they could probe dynamic correlation effects in a SSK setup.  
%\revDS{Ruggi: would you agree with my interpretation of temperaturef.} 

%%%%%%%%%%%%%%%%%%%%%%%%%%%%%%%%%%%%%%%%%%%%%%%%%%%%%%%
\section{Interpreting polaritonic chemistry with the SSK model~\label{sec:consequences}}

The similarities between the SSK model, defined in Eq.~\eqref{eq:ssk}, and the cavity-mediated electron correlations in Eq.~\eqref{eq:spinglass_gen}, are striking. In the following section, we derive different theoretical consequences of the potential spin glass nature of cavity-mediated electron correlations, and present a concise picture of its consequences for chemistry under VSC. Eventually, we briefly connect our theoretical picture to recent experimental result. %Note, however, that we here only stress similarities. Whether this picture can explain the observed effects of molecular ensembles under VSC needs extensive future theoretical, numerical and experimental efforts. 

\subsection{SSK electron correlations }

The aforementioned similarity suggests the following assignment for the total free energy of the intermolecular DSE correlations
\begin{eqnarray}
     F_{\rm corr}^{\rm DSE} \sim  N_J F_{T,\tilde{J}}.
\end{eqnarray}
This implies that we explicitly impose normally-distributed random interactions for $J_{ij}$. Realistic molecular ensembles in a cavity will certainly deviate considerably, and we expect rather heavy-tailed distributions to appear in nature, i.e., relatively few degrees of freedom contribute strongly to $J_{ij}$, whereas the others are only marginally (rare-event driven). Determining realistic distributions is non-trivial and requires considerable future research efforts. Clearly, the spin glass properties will be affected by the distribution. Nevertheless, the overall picture developed in Sec.~\ref{sec:spin_glass} should still apply, since the generic features of spin glasses remain preserved qualitatively across many known models in the literature.~\cite{parisi_nobel_2023} As an immediate consequence of collective strong coupling, thermal effects must be included when solving the dressed electronic problem, even-though the thermal energy scale can be orders of magnitude smaller than the free-space excitation of the  electronic-structure of a single molecule. This can be rationalized by the fact that the novel collective electronic excitations can be on a much lower energy scale, similar to solid-state systems. In particular, we expect a cavity-induced spin glass phase transition to occur if all prerequisites, discussed in Sec~\eqref{sec:SSKmodelemerges}, are met by the polaritonic system.
Indeed, Eq.~\eqref{eq:freefluct} suggest differently distributed (and differently scaling) fluctuations of the electronic correlations for $T<T_c$. Notice, while formally the changed fluctuation is localized per spin, this does not imply that the corresponding intermolecular DSE correlations is localized in space. Indeed, we expect that rather de-localized orbitals are affected, which contribute to the long-range intermolecular dispersion effects. Those (rare events) are expected to allow for significant overlap integrals, represented by the tails of the $J_{ij}$-distribution. Nevertheless, the proposed spin glass phase transition provides a theoretical mechanism, which could explain, why under many circumstances no chemical changes are observed, i.e., $T>T_c$. However, under specific but non-trivial conditions, $T<T_c$ can be reached, which implies that rare events and fluctuations scale differently. These changes can become chemically relevant under collective strong-coupling conditions.

Apart from the modified fluctuations in a static picture, the SSK mapping suggests unique dynamic features. In particular, the emergence of long-lived time-correlations and aging effects (see Figs.~\ref{fig:correlations} \& \ref{fig:aging}) effectively prevents the DSE electron correlation dynamics to (ever) reach thermal equilibrium even for long waiting times at $T<T_c$. This suggests a cavity-induced time-reversal symmetry breaking. This could also open novel pathways for the sub-field of chiral polaritonics.~\cite{bustamante_relevance_2024} There the common aim is to reach cavity-induced enantioselctivity by explicitly parity-violating cavity polarizations.~\cite{hubener_engineering_2021,sun_polariton_2022, schlawin_cavity_2022, voronin_single-handedness_2022, mauro_chiral_2023, baranov_toward_2023,riso_strong_2023,riso_strong_2024} However, from a fundamental theoretical aspect similar effects may also be reached by spontaneous symmetry breaking instead~\cite{greiner_field_1996,srednicki_quantum_2007} and/or degenerate states.\cite{bustamante_relevance_2024}  
At this point it is important to mention that overall our dynamic picture remains incomplete. In particular, in a polaritonic setup, the entire physical system will evolve, i.e., the nuclear and displacement field parameters $R,q_\beta$ will explicitly depend on time and thus $J_{ij}\mapsto J_{ij}(t)$, which is not considered by the discussions in  Sec.~\ref{sec:spin_glass}. In that sense, our dynamic picture of the electrons can only be considered as the limiting case of infinitely-slow dynamics of the external parameters (quasi-static). How the different involved time-scales favor or suppress the emergence of non-equilibrium aging effects remains an open question. However, the emergence of non-equilibrium electron dynamics could have fascinating consequences on the thermal equilibrium features of the entire ensemble, as we would like to briefly argue in the following.

%%%%%%%%%%%%%%%%%%%%%%%%%%%
\subsection{Non-canonical nuclear dynamics / stochastic resonances \label{sec:resonance}}
Returning to our initial Born-Oppenheimer partitioning, we notice that approximating the "ground-state" dynamics of the nuclei and displacement field classically (see Eq.~\eqref{eq:class_ham}) becomes considerably more complex.
Even when ignoring non-adiabatic couplings for the highly degenerate polaritonic ground-state, i.e., sustaining a classical picture, the non-equilibrium electron dynamics in the aging regime implies that the electronic force contributions become explicitly time-dependent within the spin glass phase. Therefore, 
%\revDS{here I am not so sure about a correct / meaningful notation. Typically one would take the gs that determines $\boldsymbol{c}$. However, we have a degenerate system now. Shall we also indicate the waiting time. However, to me it is not clear what it would mean in our context. therefore, I would ommit it here. }
\begin{eqnarray}
\mathcal{E}\big(\boldsymbol{R}(t),q_\beta(t)\big)\overset{T<T_c}{\mapsto} \mathcal{E}\big(\boldsymbol{R}(t),q_\beta(t),t\big).%= E_0\big(\boldsymbol{R}(t),q_\beta(t)\big)+ E_{\rm corr}\big(\boldsymbol{R}(t),q_\beta(t),t\big)
\end{eqnarray}
In that regard, the spin glass nature of the  electronic-structure bridges not only different length and time-scales, but it also breaks the conservative nature  of the classical forces of Eq.~\eqref{eq:class_ham}. While the magnitude of the cavity-induced non-equilibrium aging dynamics might be very small, their correlated and long-lived nature may still be sufficient to introduce stochastic resonance phenomena.~\cite{sidler_perspective_2022} 
%At least for certain chemical systems, resulting in chemically relevant effects for the thermodynamics of the nuclei. 
As we know from a set of different experiments addressing a variety of chemical reactions in cavities (see, e.g., Sec.~\ref{sec:nmr}), resonance effects (under normal incidence\cite{patrahau_direct_2024}) play a major role when modifying chemical properties under collective VSC.~\cite{thomas_groundstate_2016,thomas_tilting_2019,thomas_ground_2020} Thus it comes as no surprise that periodic feedback effects between the frustrated (off-equilibrium)  electronic-structure, the nuclear and cavity degrees of freedom and the thermal bath re-appear in a holistic theoretical description.~\cite{sidler_perspective_2022} Considerable theoretical efforts went into investigating resonance effects with polaritonic reaction-rate theories~\cite{li_cavity_2021,li_theory_2021,yang_quantum_2021,mandal_theory_2022,lindoy_resonant_2022,cao_generalized_2022,lindoy_quantum_2023,ke_quantum_2024,ke_stochastic_2024,ying_resonance_2024,montillo_vega_theory_2024} or few-molecule ab-initio simulations~\cite{schafer_shining_2022,sun_polariton_2022,philbin_chemical_2022}, but so far no consensus emerged with respect to the fundamental mechanism(s) at work.
%\revDS{how to show, what to mention, add a bit more standard points and probably link picture below to it.}, 
%we believe that the spin glass concept provides a necessary link between the macroscopic / collective (rates) and microscopic (few-molecule) chemical pictures. It opens up the door to identify prerequisites to observe chemical changes in a cavity from an ab-initio perspective. In more detail, we anticipate a scale-connecting resonance mechanism under collective VSC of a form similar to the following: Assuming that the losses of the cavity are sufficiently small (an effective cavity mode is distinguished), a dynamic instability (phase transition) of the dressed  electronic-structure can emerge for sufficiently strong collective couplings. The resulting polarization-glass phase introduces  dynamic correlations between the  electronic-structures of the molecules. The long-lived correlations imply modifications of the standard (thermal) fluctuation-dissipation relations for the dressed  electronic-structure, which implies a different out-of-equilibrium description. Eventually, the dynamic (memory-dependent) electronic correlations will act back on the dynamics of the photon displacement field and the nuclei. 
Clearly, the picture of spin glass-triggered resonance effects remains vague at the moment, and substantial theoretical and experimental effort will be required to unravel it and make it more quantitative and predictive. For the moment, we can only check if the implications of the SSK model agree qualitatively with experimental data.

%\revDS{now interpret the findings for our case, i.e. which regime? can spin glass community learn something from VSC? potential infinite waiting time limit much faster reached? Mention that instead of $T_g$ we turn on cavity. Connect to experimental equilibrium effects, after putting thins into a cavity? What about time-scales? Different noise than temperature? Mention memory and rejuvenation? Connect to numerical simulation on large computers \citenum{baity-jesi_statics-dynamics_2017}. Probably we can use cavity to better check theoretical predictions, if equilibrium is reached much faster and correlation drops quicker....How can we avoid computational complexity in VSC? also approximate chieq, with applied field and then coupling to cavity? is what we do in scf-cycles, when adding classical changes first? Here seems to be a difference between $T_g$ and $\lambda$ on / off? How can we connect to nucleation? Connect aging interpretation and rough pdf of cH eqs. Think about computational tools to converge cH equations for stronger couplings... Connect nucleation to distribution, or (?) probably better to excess of fluctuations (?). Expect local polarization to be small.}

%\begin{figure}

 %    \label{fig:off-equilibrium}
%\end{figure}

    \begin{figure}
     %\begin{subfigure}[b]
         \centering
         \includegraphics[width=0.45\textwidth]{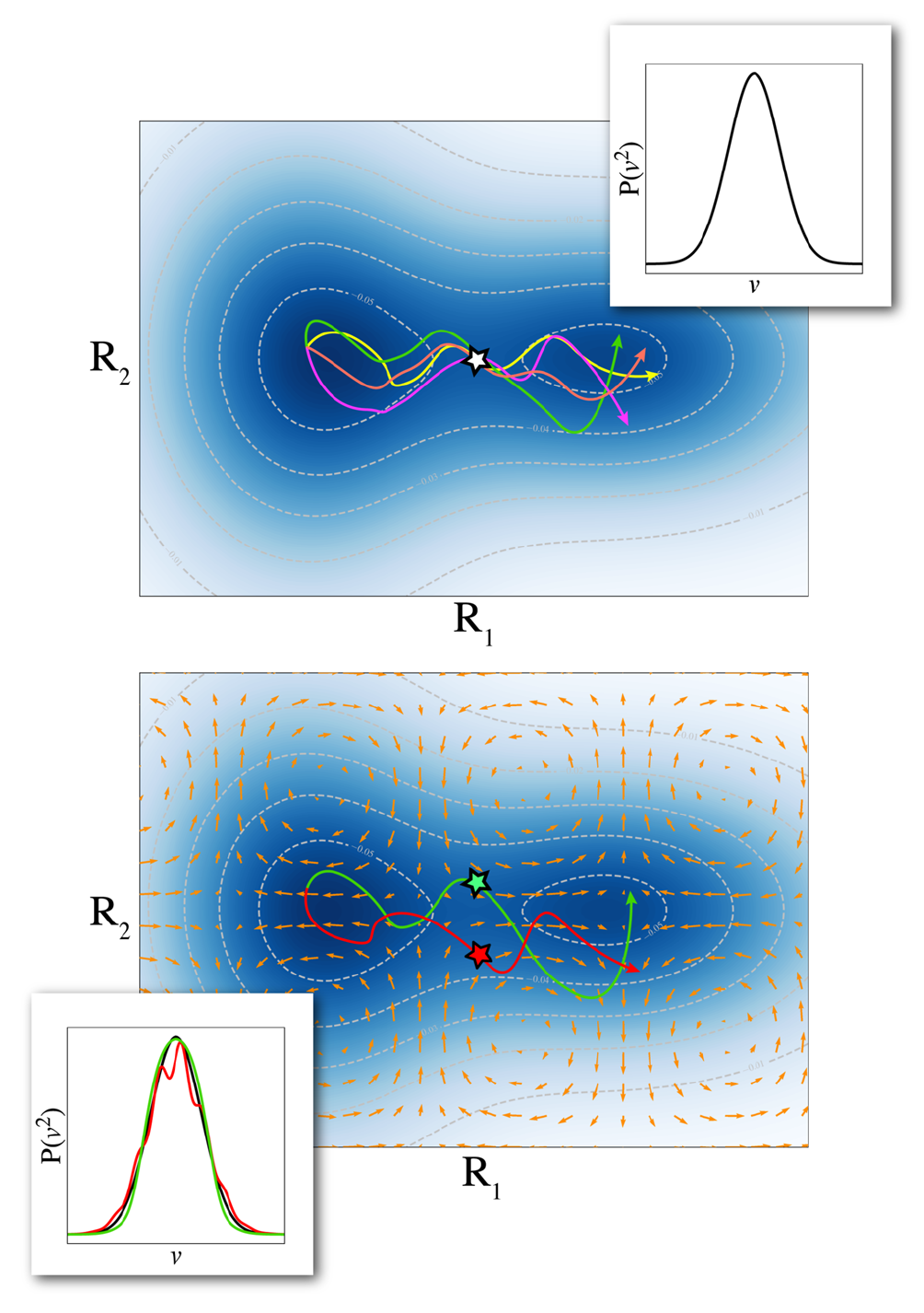}
         \caption{Illustrative sketch of classical nuclear trajectories evolving on a conservative potential energy surface (top), whereas aging effects (time-correlations) of the cavity-mediated correlation energy (bottom) may introduce non-conservative forces (orange arrows). The coupling to a thermal bath (Langevin equations), will result in canonical thermal equilibrium for the nuclei (top), whereas cavity-mediated spin glass effects are expected to give rise to (weak) non-canonical nuclear dynamics, prone to build up stochastic resonances (from Ref.\citenum{sidler_perspective_2022}).  }
         \label{fig:nmr}
     %\end{subfigure}
\end{figure}
%\revDS{discuss / mention dynamic correlation and resonance effects, feedback between nuclei and ? Can negative correlation imply negative temperature changes?}

\subsection{Experimental evidence \label{sec:nmr}}

In the previous subsections we have seen that the emergence of a cavity-induced spin glass phase transition of the  electronic-structure has several fascinating physical implications for an ensemble of molecules embedded in an optical cavity at ambient temperature $T$.
%, with one important difference: The ability of molecules to continuously polarize is violated by the two level approximation of the SK model, which eventually causes the difference  $E_{\rm cH}[\Delta\mu_0]=0$ vs. $E_{\rm SK}[\Delta\mu_0]\neq0$. At the moment, we would  consider this as the major criterion to distinguish theoretically a polarization glass from a traditional spin glass. I
%
%Moreover, finding $E_{\rm cH}[\Delta\mu_0]=0$ has important experimental implications, since it suggests that experiments that are sensitive to average changes of the molecular polarizabilities (e.g. NMR) will most likely measure no impact on the  electronic-structure under VSC.
%
In the following, we first compare the above theoretical predictions specifically with recent experimental nuclear magnetic resonance (NMR) results under VSC.~\cite{patrahau_direct_2024} In Figs.~\ref{fig:nmr}~a-c the influence of VSC on the equilibrium concentration between two conformations of a molecular balance, sensitive to London dispersion forces, is studied with NMR spectroscopy. The experiments reveal that VSC can indeed modify the equilibrium rate constant and thus changes chemical properties that are directly related to electron-electron correlations, provided that the cavity is tuned on resonance with a specific C-H stretching mode of the solute molecules. 
%\revDS{how should we relate the resonance feature to our previous discussion?} 
If we disregard the resonance feature for the moment, the following three experimental observations seem to directly relate to our proposed spin/polarization-glass mechanism: First, the absence of cavity-induced chemical shifts indicates that the  electronic-structure is \textit{on average} not polarized by the cavity, which perfectly agrees with what we would anticipate from Eq.~\eqref{eq:ch_mean}, i.e., the polarization glass. Second, the broadening of the chemical shifts seems modified under VSC, which one would expect for a cavity-induced polarization glass, i.e., from Eq.~\eqref{eq:ch_var}, or for modified rare events, i.e., for modified distribution tails that give rise to a spin glass phase. Notice, however, that in Ref.~\citenum{patrahau_direct_2024} the modified broadening is assigned to experimental artifacts and thus not further investigated. The third important insight of the NMR experiments is an abrupt change of the equilibrium constant at a specific collective strong coupling strength (influenced by the concentration) that does not scale further with the collective Rabi splitting. This suggests a phase transition at a specific collective coupling, as suggested by the SSK model. Notice further that while collective strong coupling (Rabi-splitting) has been reached in (dilute) gas phase experiments, so far no (local) change in chemistry was observed.\cite{wright_rovibrational_2023,wright_versatile_2023,nelson_more_2024} This indicates that more condensed molecular ensembles (which possess significant dispersion effects) may be an crucial ingredient to change chemistry and thus modify matter-properties locally. This observation is inline with what we can derive from the quasi-dilute gas picture, which is an essential prerequisite for the formation of spin glass-like correlation effects of the  electronic-structure. 

In the following, we briefly look at a series of further experimental results that nicely connect, with what one would expect qualitatively from our mapping of the DSE electron correlations onto the SSK model. 
\begin{enumerate}
    \item  Cavity-modified \textit{dispersion forces} are known to play a crucial role in a series of different experiments, which report chemical changes under collective strong coupling.  In particular, various cases have been reported where cavity-modified London-type dispersion forces\cite{patrahau_direct_2024} alter the self-assembly of molecular structures under collective (and cooperative) VSC\cite{joseph_supramolecular_2021,sandeep_manipulating_2022,joseph_consequences_2024}. This macroscopic ordering across the molecular ensemble nicely illustrates the long-range nature of the all-to-all intermolecular DSE interactions.

    \item Two different cases of \textit{phase-like transitions} have been reported with respect to the collective coupling strength. For certain experiments (e.g., charge-transfer complexation\cite{pang_role_2020}, supramolecular assembly of conjugated polymers\cite{joseph_supramolecular_2021}, conformational equilibrium constants\cite{patrahau_direct_2024}), an abrupt qualitative change is observed, with little dependency on the collective coupling strength beyond the critical point, whereas other cases seem to undergo a phase transition, with continuing parametric dependency (e.g. reaction rates\cite{thomas_groundstate_2016} or conductivity measurements in Refs. \citenum{kaur_controlling_2023,kumar_extraordinary_2024}). Notice, that typically the transition is not directly connected to the emergence of a Rabi splitting.\cite{thomas_ground_2020} Having in mind the abrupt change of the probability distribution of the free energy fluctuations of the SSK model, given in Eq. \eqref{eq:freefluct}, different transition regimes must be anticipated. In particular, observables that rather depend on the mean of the distribution (i.e., \textit{bulk properties} such as  conformational equilibrium or self-assembly\cite{pang_role_2020,joseph_supramolecular_2021,patrahau_direct_2024}) are expected to show only a weak  parametric dependency. In contrast, \textit{rare events} (e.g. chemical reactions,\cite{thomas_groundstate_2016} tunneling events\cite{kumar_extraordinary_2024}), that can rather be considered a measure for the tails of the probability distribution, are indeed sensitive to small changes of the collective coupling strength beyond the critical point. Remark: A phase-like transition on aggregation properties can also be observed for ground-state properties under electronic strong coupling. This suggests that our spin glass picture is transferable to electronic strong coupling situations.\cite{biswas_electronic_2024} However, this aspect goes beyond the scope of this work. 

    \item  Minimizing the DSE correlation energy (i.e., approaching the collective ground state) in Eq. \eqref{eq:spinglass_gen} favors the \textit{delocalization} of the intermolecular electronic orbitals. We anticipate that this effect should overall enhance tunneling effects\cite{kumar_extraordinary_2024} and increase the conductivity\cite{jarc_cavity-mediated_2023} or reactivity of the chemical ensemble under strong coupling along the polarization axis of the cavity. However, the impact of the DSE delocalization on chemical properties orthogonal to the polarization axis is difficult to anticipate.  Depending on the chemical system, dispersion interactions may effectively be increased or suppressed when considering them spatially averaged. This could explain why it is delicate to control rare events, i.e., chemical reactions may either be suppressed\cite{thomas_groundstate_2016,hirai_modulation_2020} or increased.\cite{lather_cavity_2019} In that regard the symmetry properties of light and matter may play an important role and could help to make qualitative predictions.\cite{sandeep_manipulating_2022}  

        \item The inherent non-equilibrium nature of the SSK dynamics with no replica symmetry breaking suppresses many typical spin glass features, which makes the verification of our theoretical hypothesis particularly hard. Still, the trivially-modified fluctuation dissipation relations in Eq. \eqref{eq:modflucdis}  suggest an overall increase of the fluctuations (\textit{heating}) in the aging regime and the emergence of  \textit{cavity-modified hysteresis} effects. Both effects have been reported in various experiments: For example, an effective temperature increase was measured for melting temperature of dsDNA,\cite{zhong_driving_2023}, supramolecular polymerization\cite{joseph_supramolecular_2021} or metal-to-insulator transitions in 1T-TaS$_2$\cite{jarc_cavity-mediated_2023} and for the large enhancement of ferromagnetism under VSC.\cite{thomas_large_2021} Cavity-modified hysteresis is for example shown in Refs.\citenum{thomas_large_2021,jarc_cavity-mediated_2023}. Notice, from a theoretical point of view, it may be surprising that an increase of the effective temperature can be accompanied by an ordering (self-assembly) of the system. However, as recently shown and explained generically in Ref.\cite{morone_re-entrant_2024}, such an effect is characteristic for many complex system.
\end{enumerate}

At this point it is important to note that the experimental evidence for our spin glass mapping remains vague and each of the mentioned aspect needs considerable research effort to be validated or falsified. In particular, the role of (stochastic) resonances and multiple time-scales remains unknown. Nevertheless, to our opinion many of the reported experimental results are in excellent qualitative agreement, with what one would anticipate from the physics of the SSK model.  
%
%\revDS{what other experimental evidence should be mentioned and how? Fausti? We should be specific. The modified fluctuations / off-equilibrium aspects should also be mentioned...}
%The numerical simulations of the Shin-Metiu molecule have (so far) not unraveled a resonance conditions. We anticipate that this experimental simple aspect is highly non-trivial to capture theoretically within the cBO setup, since it must include feedback effects from the nuclear dynamics acting on the polarization glass, which likely require very long simulation times to emerge from first principles.
%We will come back to this important aspect at the end of this review (see Sec.~\ref{sec:offequil}), where we will argue that the resonance mechanism is likely a manifestation of the dynamical aspect of the cavity-induced frustration mechanism. 
\begin{figure}
     %\begin{subfigure}[b]{0.45\textwidth}
         \centering
         \includegraphics[width=\textwidth]{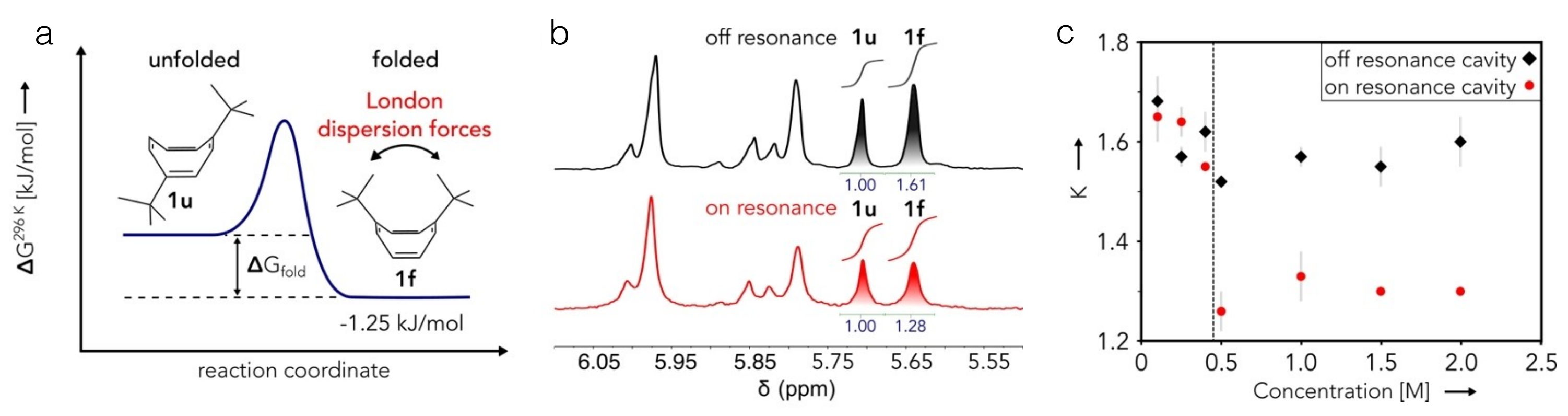}
         \caption{Influence of VSC on the London dispersion-force-driven equilibrium, determined from NMR measurements according to Ref.~\citenum{patrahau_direct_2024}: (a) Free energy difference between folded (1f) and unfolded (1u) conformer in benzened-$d_6$. (b) On- (red) and off-resonance (black) $^1$H NMR spectra of a 1 M solution. The shaded peaks allow to distinguish the two different conformers. The absence of a frequency shift between the two spectra indicates that on average the  electronic-structure is not affected (no chemical shift). The magnitude of the peaks can be related to the equilibrium constant between the two conformers, which indicates VSC-induced chemical changes. (c) Varying the concentration (i.e., the collective coupling strength) reveals a critical concentration (dotted line), where suddenly a different conformational equilibrium constant is approached, which suggests the emergence of a cavity-induced phase transition.  }
         \label{fig:nmr}
     %\end{subfigure}
\end{figure}
Therefore, we believe that the spin glass-like nature of the dressed  electronic-structure provides the most realistic and plausible theoretical framework that sets the necessary seed (instability mechanism) to trigger the resonance effects that have been observed in polaritonic experiments. Consequently, we consider the connection between polaritonic chemistry and the physics of a spin glass an excellent starting point to not only better understand current experimental findings, but also to stimulate novel theoretical and experimental directions. This should allow us to reach a holistic understanding of strongly-coupled light-matter systems. Moreover, our connection provides an interesting new direction for the field of spin glasses beyond condensed matter systems.  

\section{Summary and Conclusion~\label{sec:conclusion}}

In this work, we highlight a novel and fundamental connection between two previously unrelated research domains: the emerging field of polaritonic chemistry and the well-established field of spin glasses and rare events in statistical mechanics. By mapping the cavity-mediated electron correlations in a quasi-dilute molecular ensemble under collective vibrational strong coupling (VSC) onto the spherical Sherrington–Kirkpatrick (SSK) model of a spin glass, our results suggest the emergence of a cavity-induced phase transition.
%In this work we have highlighted a novel connection between two hitherto separate research areas: The emergent field of polaritonic chemistry and the well-studied field of spin glasses and rare events in statistical mechanics.
%The mapping of the (cavity-mediated) electron correlations of a quasi-dilute molecular ensemble under collective vibrational strong coupling (VSC), onto the spherical Sherrington-Kirkpatrick (SSK) model of a spin glass, suggests the emergence of a cavity-induced phase transition. 
\\

The theoretical pre-requisites for the proposed cavity-mediated spin glass phase can be summarized as follows:
\begin{enumerate}
    \item Polarization ordering (a polarization-glass phase) emerges for sufficiently strong collective light-matter couplings to the cavity modes in the dilute gas limit. Resulting numerical convergence issues indicate a \textit{collectively-degenerate electronic ground-state} for the molecular ensemble under VSC.
    \item In the \textit{quasi-dilute gas limit}, dipole self-energy(DSE)-modified electron-electron correlations $E_{\rm corr}^{\rm DSE}$ can be derived using the configuration interaction method, which dominate over the intermolecular Coulomb correlations. In that case, the intermolecular DSE-correlation energy has the form of a spin glass, provided that the density of states at the ground-state energy is sufficiently large. In particular, the DSE interactions are considered random due to the random orientation of the molecular ensemble with respect to the polarization of the cavity.
    \item The DSE correlations connect to the SSK model when assuming $E_{\rm corr}^{\rm DSE}\sim H_{SSK}$. This suggests a \textit{spin glass-like phase transition} induced by the cavity, provided that the fluctuations of the random DSE interaction are sufficiently strong.  In particular, those fluctuations must be stronger than the thermal fluctuations of the dressed electronic correlations, i.e. $T<T_c$.   
\end{enumerate}

Entering a cavity-mediated spin glass phase using the SSK model suggests the following theoretical \textbf{consequences} for our polaritonic ensemble and polaritonic chemistry phenomena in general, which are in line with various experimental results (see Sec. \ref{sec:nmr}): 
\begin{enumerate}
    \item The emergence of a spin glass-like phase transition, with respect to the collective light-matter coupling, confirms the relevance of cavity-mediated intermolecular dispersion effects. Moreover, the abrupt change of the local free-energy fluctuations occurs with respect to the number of collectively coupled degrees of freedom. This can be interpreted as a change in the probability distribution of "local" dispersion effects, which suggests a qualitatively different behavior for bulk properties (e.g., self-assembly) and rare events (e.g., tunneling, reactivity) with respect to the collective coupling strength under VSC. Overall, the minimization of the DSE correlation energy favors the delocalization of the intermolecular orbitals along the polarization axis. However, perpendicular consequences (and symmetry in general\cite{sandeep_manipulating_2022}) must not be discarded to reach for a holistic chemical picture.  
    %\item The \textit{phase transition of the local free-energy fluctuations} occurs with respect to the number of collectively coupled degrees of freedom $N_J$. This not only demonstrates locally modified fluctuations, but the assigned scaling feature could also provide a method for the experimental verification and characterization of the spin glass picture, as well as to determine the, typically unknown, quantity $N_J$.
    \item Thermal and non-adiabatic effects start to play a role for the collective electronic correlation problem in a cavity, even though these effects can usually be disregarded for an ensemble in free space. In particular, the SSK model suggests (extremely) \textit{long-lived time-correlations} that effectively prevent the cavity-mediated electron correlations to reach equilibrium. As a consequence, an overall heating as well as changes of hysteresis effects can be anticipated from the SSK model, which is in line with experimental observations.
    \item The non-equilibrium  dynamics of the cavity-mediated  electronic-structure implies non-conservative nuclear forces when assuming classical dynamics. Consequently, when coupling the classical degrees of freedom to a thermal bath (weakly-coupled Langevin dynamics), canonical equilibrium is no longer reached. Furthermore, non-conservative forces favor the build up of \textit{stochastic resonances}. While a strong resonance behavior is known experimentally since the earliest experiments,\cite{thomas_groundstate_2016} considerable research effort will be required to disentangle the consequences of our spin glass picture on the formation of resonances under collective VSC.  
     %The implied break down of standard thermal fluctuation-dissipation relations in spin glasses suggests that most physically resolvable processes are determined by frustrated off-equilibrium phenomena, which e.g. can be re-interpreted as an effective change in temperature.
\end{enumerate}

Overall the SSK model suggests the existence of an instability of the dressed electronic subsystem, altering the established temporal and spatial scales to understand chemistry. Nevertheless, the interplay of the different scales (e.g. waiting time vs. stationary correlation  vs. aging regime vs. inverse cavity frequency, ...) still remains unexplored. The polarization spin glass picture can be considered as the limiting case of infinitely
slow dynamics ($R, q_\beta$ ) (quasi-static picture). Overcoming this limitation will require further theoretical and computational efforts. In addition, determining chemically more realistic (heavily-tailed) probability distributions for the random interactions $J_{ij}$ will be challenging but important as well. The physical properties (e.g. scaling) of the resulting SSK model may deviate considerably from the case of a normally-distributed SSK model and it will be non-trivial to determine, even if realistic probability distributions are known. To gain a detailed understanding of all of the above aspects will require very interdisciplinary research efforts. However, we believe that pursuing this path will be extremely fruitful in many aspects. It may not only provide the missing theoretical piece to unravel the mysteries of polaritonic chemistry,\cite{schwartz_erc_2024} but it could also trigger novel fundamental findings. For example, it could provide an theoretical and experimental tool to better understand and quantify rare events and nucleation (via the tails of $J_{ij}$). As we all know, chemistry, e.g., a chemical reaction, is an inherently rare off-equilibrium process, which is hard to describe apart from its rate. We therefore anticipate that cavity-controlled changes of the fluctuation-dissipation relations may eventually not only help to understand polaritonic chemistry, but they could provide novel insights into rare event and nucleation processes in general.

 \section*{Biographies}

 \textbf{Dominik Sidler}, born in Ebikon, Switzerland, is a group leader at the Max-Planck Institute for the Structure and Dynamics of Matter (MPSD) in Hamburg, Germany. He is also a Scientist at Paul Scherrer Institute (PSI) in Villigen, Switzerland. He obtained his Ph.D. in computational chemistry from ETH Z\"urich, Switzerland, in 2018. Afterwards, he moved as a postdoctoral fellow to the MPSD in Hamburg, Germany. In 2024, he returned as a Scientist to PSI Switzerland and was appointed research group leader at MPSD. His research group works on the theoretical and computational description of (strong) light-matter interactions in polaritonic chemistry. He is a team member of the international ERC Synergy project "UnMySt" and  his current research focus is on first-principles approaches to describe emergent local and collective strong coupling phenomena in optical cavities, as well as the impact of vacuum field fluctuations on molecular non-equilibrium conditions at finite temperatures.

\textbf{Michael Ruggenthaler} is research group leader at the Max-Planck Institute for the Structure and Dynamics of Matter in Hamburg, Germany. He obtained his Ph.D. in theoretical physics jointly from the Max-Planck Institute for Nuclear Physics and the University of Heidelberg, Germany, in 2009. As Schr\"odinger Fellow of the Austrian Science Fund he worked at the University of Jyv\"askyl\"a, Finland, and then established an independent research group at the University of Innsbruck, Austria, before he moved to the Hamburg in 2016. His research group works on the theoretical and mathematical foundations of quantum many-body theories, in and out of equilibrium. He has been involved in several successful international research collaborations such as the QuantEra project "RouTe" or the ERC Synergy project "UnMySt". His current research focus is on first-principles approaches to ab initio quantum electrodynamics, polaritonic chemistry and cavity-modified material properties.

\textbf{Angel Rubio}, born in Oviedo, Spain, is the Director of the Theory Department of the Max Planck Institute for the Structure and Dynamics of Matter. He is also  Distinguish Research Scientist at the Simons Foundation’s Flatiron Institute (NY, USA) and Professor/Chair for condensed matter physics at the University of the Basque Country in Donostia-San Sebastián, Spain.  He received his PhD in Physics with honors from the University of Valladolid (UVA) in 1991 where he did fundamental work on the structural and optical properties of metallic clusters. Then moved to a postdoctoral position at UC Berkeley-Physics (92-95) where he predicted a new type of boron-nitride nanotubes triggering their ensuing experimental synthesis. Between 1994 and 2001 as Professor at UVA   he started the ab initio materials research open-source project octopus used now by over 1000 groups worldwide. Diverse Professorships at École Polytechnique Paris, FU Berlin and Montpellier followed. In 2001 he moved as Chair of Condensed Matter Physics at UPV/EHU in San Sebastian. There he engaged in highly successful work on modeling of excited-state properties of materials and nanostructures setting the foundations of modern theoretical spectroscopy. In August 2014 he accepted the position as Max Planck Director. There he has pioneered the development of quantum electrodynamical density functional theory (QEDFT), a novel theoretical framework for strong light-matter phenomena in chemistry and materials sciences . His work has been recognized by several awards, including the 2023 Spanish National Physics Prize “Blas Cabrera” 2018 Max Born medal and prize, 2016 Medal of the Spanish Royal Physical Society and the 2014 Premio Rey Jaime I for basic research, and more, and elected member of different academies, including the German Leopoldina Academy and Berlin-Brandenburgischen Akademie der Wissenschaften, the European Academy of Sciences, the Academia Europaea, and a foreign associate member of the National Academy of Sciences (USA).

\begin{acknowledgement}
We gratefully acknowledge all members of our ERC synergy grant proposal team (Tal Schwartz, Thomas Ebbesen, Abraham Nitzan, Sharly Fleischer, Cyriaque Genet and Maxim Sukharev) for inspiring discussions and helpful comments. Furthermore, we particularly thank Benjamin Lev, Markus Mueller, Dries Sels and Antoine Georges for their valuable and critical feedback on the spin glass aspects. Finally, we would like to thank all four referees for their very constructive feedback. Specifically, the comments of referee 3 were extremely helpful to connect our predictions to experimental observations. 
This work was made possible through the support of the European Research Council (ERC-2024-SYG-101167294, UnMySt), the Cluster of Excellence Advanced Imaging of Matter (AIM), Grupos Consolidados (IT1249-19) and SFB925.   We acknowledge support from the Max Planck-New York City Center for Non-Equilibrium Quantum Phenomena. The Flatiron Institute is a division of the Simons Foundation.
\end{acknowledgement}

%\begin{suppinfo}

%\end{suppinfo}
\bibliography{revision}
\end{document}